\documentclass[a4paper,12pt]{article}
\usepackage{amsmath}
\usepackage{amssymb}
\usepackage{fullpage}
\usepackage{verbatim}

\usepackage[T1,OT1]{fontenc}

\usepackage{graphicx}

\allowdisplaybreaks[3]

\usepackage[hang]{footmisc}
\usepackage[compress,square,numbers]{natbib}
\usepackage[colorlinks,linktocpage,linkcolor=blue,citecolor=blue,urlcolor=blue]{hyperref}

\renewcommand{\d}{\mathrm{d}}
\newcommand{\e}{\mathrm{e}}
\newcommand{\w}{\wedge}

\newcommand{\nl}{\notag \\ &\quad\,}
\newcommand{\nll}{\notag \\ &}

\begin{document}

\numberwithin{equation}{section}

\thispagestyle{empty}

\begin{flushright}
\small LMU-ASC 14/16\\
\normalsize
\end{flushright}
\vspace*{2cm}

\begin{center}

{\LARGE \bf Tachyons in Classical de Sitter Vacua}

\vspace{2cm}
{\large Daniel Junghans}\\

\vspace{1cm}
Arnold-Sommerfeld-Center f{\"{u}}r Theoretische Physik\\
Department f{\"{u}}r Physik, Ludwig-Maximilians-Universit{\"{a}}t M{\"{u}}nchen\\
Theresienstra\ss e 37, 80333 M{\"{u}}nchen, Germany

\vspace{1cm}
{\upshape\ttfamily daniel.junghans@lmu.de}\\

\vspace{2cm}

\begin{abstract}
We revisit the possibility of de Sitter vacua and slow-roll inflation in type II string theory at the level of the classical two-derivative supergravity approximation. Previous attempts at explicit constructions were plagued by ubiquitous tachyons with a large $\eta$ parameter whose origin has not been fully understood so far. In this paper, we determine and explain the tachyons in two setups that are known to admit unstable dS critical points: an SU(3) structure compactification of massive type IIA with O6-planes and an SU(2) structure compactification of type IIB with O5/O7-planes. We explicitly show that the tachyons are always close to, but never fully aligned with the sgoldstino direction in the considered examples and argue that this behavior is explained by a generalized version of a no-go theorem by Covi et al, which holds in the presence of large mixing in the mass matrix between the sgoldstino and the orthogonal moduli. This observation may also provide a useful stability criterion for general dS vacua in supergravity and string theory.
\end{abstract}

\end{center}

\newpage

\section{Introduction}

The construction of de Sitter (dS) vacua in string theory has received a lot of attention in recent years, motivated mainly by cosmology but also by more conceptual questions such as holography. Although a number of interesting scenarios and models have been proposed for how such vacua can in principle arise in string theory (some even with semi-realistic particle physics \cite{Cicoli:2013cha}), it is difficult to construct fully explicit microscopic solutions that realize these models at the level of the 10D or 11D equations of motion. One reason for this is that all candidate constructions have to be rather elaborate in order to evade certain no-go theorems that forbid dS vacua in the simplest supergravity compactifications \cite{Gibbons:1984kp, deWit:1986xg, Maldacena:2000mw}. An additional complication is the requirement of moduli stabilization, which, in the absence of supersymmetry, is expected to be achieved only by a small fraction of the critical points in moduli space. Most string theory constructions overcome these hurdles at the cost of including (among other ingredients) non-perturbative quantum corrections to the 4D scalar potential, which are not well understood at the 10D level (see, however, \cite{Koerber:2007xk, Baumann:2010sx, Heidenreich:2010ad, Dymarsky:2010mf} for some interesting results). Examples for constructions of this type include the KKLT scenario \cite{Kachru:2003aw, Bergshoeff:2015jxa}, the large volume scenario \cite{Balasubramanian:2005zx, Conlon:2005ki}, the K{\"{a}}hler uplifting scenario \cite{Balasubramanian:2004uy, Westphal:2006tn, Rummel:2011cd, Louis:2012nb} as well as a number of more recent proposals \cite{Cicoli:2012fh, Blaback:2013qza, Danielsson:2013rza, Rummel:2014raa, Kallosh:2014oja, Marsh:2014nla, Guarino:2015gos, Retolaza:2015nvh}. On top of that, in some (but not all) constructions, uplifting terms have to be added to the scalar potential in order to make the vacuum energy positive. Another possibility to obtain dS vacua is to include non-geometric fluxes \cite{deCarlos:2009fq, deCarlos:2009qm, Dibitetto:2010rg, Danielsson:2012by, Blaback:2013ht, Damian:2013dq, Damian:2013dwa, Hassler:2014mla}. However, in spite of promising developments (see, e.g., the recent works \cite{Danielsson:2015tsa, Blumenhagen:2015zma, Blumenhagen:2015lta} and references therein), their uplift to string theory has not been sufficiently understood so far.

In other words, it seems that complexity is the price to pay to get dS vacua in string theory, which makes it hard to spell out all details of the constructions in full explicitness. With the above applications in mind, however, this would certainly be desirable. It should therefore be useful to identify the corners of the string landscape with the simplest possible examples for dS solutions.

As an alternative to the above approaches, the idea of so-called classical dS vacua was put forward a few years ago \cite{Hertzberg:2007wc, Silverstein:2007ac}. These are vacua that are constructed purely at the level of classical two-derivative supergravity, supplemented by the presence of branes and O-planes. The initial hope was that, in contrast to constructions involving quantum corrections, models of this type might be simple enough to allow fully explicit 10D solutions. However, constructing such solutions turns out to be complicated by a number of additional no-go theorems that are more restrictive than the older theorems of \cite{Gibbons:1984kp, deWit:1986xg, Maldacena:2000mw}. These refined no-go theorems exclude dS vacua (and, more generally, slow-roll inflation) for a large class of compactifications of type IIA and type IIB string theory \cite{Hertzberg:2007wc, Haque:2008jz, Steinhardt:2008nk, Caviezel:2008tf, Flauger:2008ad, Danielsson:2009ff, Caviezel:2009tu, Wrase:2010ew, VanRiet:2011yc} (see also \cite{Gautason:2012tb, Green:2011cn, Kutasov:2015eba, Quigley:2015jia} for no-go theorems in heterotic string theory). If one restricts to the simplest case of geometric compactifications with ``non-exotic'' sources (i.e., D-branes and O-planes)\footnote{See also \cite{Andriot:2010ju, Saltman:2004jh, Dong:2010pm, Dodelson:2013iba} for other approaches to obtain dS solutions.}, a minimal requirement to evade the no-go theorems is to include O-planes and RR fluxes as well as a negatively curved internal space.\footnote{It was argued in \cite{Dasgupta:2014pma} that an even more restrictive no-go theorem holds in type IIB string theory which fully excludes dS vacua at the classical level. However, the theorem of \cite{Dasgupta:2014pma} can be evaded in the presence of O-planes such that it is essentially equivalent to the older theorem of \cite{Maldacena:2000mw} (see Appendix \ref{app} for a discussion of this point).}

Unfortunately, none of the known solutions obtained from a consistent truncation of the 10D fields yield any meta-stable dS vacua but only dS critical points with at least one tachyonic direction in moduli space \cite{Caviezel:2008tf, Flauger:2008ad, Danielsson:2009ff, Caviezel:2009tu, Danielsson:2010bc, Danielsson:2011au}. The $\eta$ parameter is always large in these solutions such that they are not suitable for slow-roll inflation either (see also \cite{Gur-Ari:2013sba, Blaback:2013fca}). Moreover, the origin of the tachyons does not seem to be explained by general no-go theorems against stability \cite{Covi:2008ea, Shiu:2011zt} such that it is a priori not clear whether they are just coincidental or due to a hidden structure shared by all of the constructions.
In \cite{Covi:2008ea}, it was argued from a 4D supergravity perspective that tachyons are expected to appear along the sgoldstino direction if the scale of supersymmetry breaking at a dS critical point is small compared to the scale of the masses of the orthogonal moduli. In \cite{Shiu:2011zt}, it was analyzed how the scalar potential of type II flux compactifications depends on the volume and dilaton moduli, and necessary criteria for stability in the volume-dilaton plane of the moduli space were worked out. However, the ubiquitous tachyons mentioned above do in general neither lie along the sgoldstino direction nor inside the volume-dilaton plane.

Important progress on this issue was made in \cite{Danielsson:2012et} where the authors analyzed a class of $\mathrm{SU}(3)$ structure compactifications with O6-planes in massive type IIA supergravity, which are known to admit unstable dS extrema \cite{Caviezel:2008tf, Danielsson:2010bc}. They then conjectured that the tachyonic directions should lie in the subspace spanned by the dilaton, the overall volume modulus and the volume moduli of the cycles wrapped by the O6-planes, which was numerically verified for various dS critical points in this class of models. Furthermore, the authors observed that the tachyons align with the sgoldstino in a specific singular limit where the dS solutions approach a supersymmetric Minkowski point and certain cycles blow up or shrink to zero size.

Nonetheless, several open questions remain. First, it is not obvious how to generalize the result to other classes of compactifications in type IIA and type IIB that admit dS critical points. Are the tachyons always a combination of the dilaton, the 6D volume and the O6-plane volumes? What about O-planes of different co-dimension? What would be the type IIB version of the rule? Second, it would be useful to identify the tachyonic directions analytically in order to gain a better understanding of their origin. This should shed light on the question \emph{why} the tachyons appear along certain directions and not along others inside the moduli space. It should also reveal why the tachyons are in general not explained by the sgoldstino argument of \cite{Covi:2008ea}. Third, one would ideally like to have a concrete recipe for how to avoid tachyons in future constructions, e.g., by specifying a list of minimal ingredients or determining promising corners in the string landscape.

The purpose of the present paper is to elaborate on the first two of these points and, based on these results, make (tentative) suggestions for how to address the third one. We will study two different models admitting families of dS critical points, one in type IIA and one in type IIB string theory. The first model is an SU$(3)$ structure compactification of massive type IIA with O6-planes and NSNS and RR fluxes, which was studied in \cite{Danielsson:2010bc} (see also \cite{Caviezel:2008tf, Danielsson:2011au, Danielsson:2012et}). Due to an isotropy property, the model is remarkably simple and therefore perfectly suited for studying questions such as those we are interested in. The second model is an SU$(2)$ structure compactification of type IIB with O5/O7-planes and RR fluxes, which was studied in \cite{Caviezel:2009tu}. The internal spaces in both cases are (different) orientifolds of the group space $\textrm{SU}(2) \times \textrm{SU}(2)$. Our main result is that, in the dS solutions we consider, the ubiquitous tachyons can be explained by a combination of two effects:

\begin{itemize}
 \item \emph{Strong metric deformations}: In order to achieve a negatively curved internal space, the internal metric is strongly deformed in all dS solutions we analyze, which brings the solutions close to a singular point in moduli space in which the scale of SUSY breaking is small compared to the moduli masses. As stated above, it was already observed in \cite{Danielsson:2012et} that the type IIA solutions degenerate to a SUSY Minkowski point in a certain singular limit. Here, we will make this observation more precise by presenting the analytic solutions, which were not known before, and by stating explicitly how they depend on a small SUSY-breaking parameter $\epsilon$. We will furthermore argue that the existence of this small parameter is not a coincidence but necessary in this class of solutions due to the requirement of negative internal curvature. Finally, we will show that analogous observations hold in the type IIB solution we study.
 \item \emph{Generalized sgoldstino theorem}: Since SUSY breaking is small in all dS solutions we discuss, the tachyons can be explained by a generalized version of the sgoldstino no-go theorem of \cite{Covi:2008ea}. As we will explicitly show, the region in moduli space surrounding the singular point exhibits large mixing between the sgoldstino and the other moduli in both classes of solutions. This has the effect that the tachyon rotates away from the sgoldstino direction as one moves away from the singular point. The generalized theorem thus predicts that a tachyon will occur in a direction close to, but never completely aligned with the sgoldstino. We will make this statement precise and analytically determine the exact tachyon directions for the family of dS solutions in type IIA. We will then show that our explanation is also correct for the type IIB dS solution of \cite{Caviezel:2009tu}. We will furthermore make contact with the proposal of \cite{Danielsson:2012et} by verifying that the instability at this critical point appears in the moduli subspace of the O5/O7-plane volumes.
\end{itemize}
We believe that our results yield a satisfactory explanation for the appearance of the tachyons in classical dS solutions, thereby shedding light on a persistent puzzle in the literature. Moreover, our observations regarding the sgoldstino may be useful more generally as a stability criterion for dS vacua in supergravity and string theory and thus facilitate a systematic scan for cosmologically interesting corners of the landscape.

Another useful perspective on tachyons is to consider statistical arguments \cite{Marsh:2011aa, Chen:2011ac, Bachlechner:2012at}.
The main idea of this approach is to randomize certain parameters of the 4D scalar potential that encode the properties of specific solutions.
Thus, one can hope to capture the essential features of the vacuum distribution in a certain region of the string landscape without having to analyze in detail all possible solutions in that region. Along these lines, it was shown in \cite{Marsh:2011aa, Chen:2011ac} that the probability of a dS critical point to be meta-stable is exponentially suppressed with some power of the number of the moduli. Since the latter is typically of the order $\mathcal{O}(10)$ even in the simplest models, this provides a reasonable explanation for why no meta-stable classical dS vacua have been found so far.

However, although statistical analyses are certainly useful to estimate the properties of the landscape, they do (by construction) not take into account the full structure imposed on the 4D scalar potential by the underlying string theory solutions. One therefore potentially misses patterns in the values taken by the parameters, which may arise, e.g., due to certain constraints or symmetries. Not taking into account such information can affect the estimated number of meta-stable vacua in a given region of the landscape. One therefore has to go beyond landscape statistics if one wants to understand whether the tachyons are just due to technical difficulties (i.e., identifying the small fraction of meta-stable vacua among the total number of critical points) or whether there are in fact structural reasons for them. In this paper, we find evidence for the latter possibility. This is useful since one may now increase the chance of finding dS solutions by performing a systematic search in corners of the landscape with an increased probability of meta-stability.

Aside from the tachyon problem, other issues have been pointed out with attempts at constructing classical de Sitter vacua. One criticism is that sources such as O-planes are usually smeared over the compact space in order to simplify the equations of motion. Since O-planes are involutions and thus intrinsically localized, a smeared solution is not a good approximation of the true solution \emph{pointwise} in the compact space. Nevertheless, at least in some regions of the moduli space (in particular, in the large-volume limit), one can hope that the smeared approximation is trustworthy for 4D objects such as the scalar potential and the cosmological constant, which are obtained by integrating certain combinations of 10D fields over the compact space. It is in this sense that the solutions discussed in this paper can be useful.\footnote{Nevertheless, promoting smeared solutions to localized ones can involve a variety of subtle effects. This includes corrections to the potential and the definition of the 4D fields (see, e.g., the recent works \cite{Martucci:2014ska, Grimm:2014efa} and references therein), the appearance of certain singularities (see, e.g., \cite{Blaback:2014tfa, Bena:2014jaa, Hartnett:2015oda, Michel:2014lva} for a sample of recent works), and sometimes even a change in the amount of preserved supersymmetries (e.g., the non-SUSY smeared solution of \cite{Blaback:2011nz} has both SUSY and non-SUSY localized counterparts \cite{Apruzzi:2013yva, Junghans:2014wda}, see also \cite{Danielsson:2013qfa}).}
Another worry is related to the unclear fate of O$6$-planes in massive type IIA supergravity \cite{Banks:2006hg, McOrist:2012yc}, where no M-theory description is available.
Some interesting results regarding this issue were obtained in \cite{Saracco:2012wc} (see, however, \cite{Gautason:2015tig}). Finally, it was pointed out in \cite{Danielsson:2011au} that it is difficult to satisfy flux quantization conditions in certain models of classical dS vacua. These issues, while clearly important, are not addressed in this paper. Instead, we will focus on the tachyon problem, which we believe is the most pressing.

This paper is organized as follows. In Section \ref{su3-IIA}, we first construct the analytic solution for the family of dS critical points in the type IIA model. We then explain the origin of the tachyon in terms of the sgoldstino argument. In Section \ref{su2-IIB}, we show that the same argument can be used to explain the tachyon in the type IIB model, and we discuss the relation to the O-plane volume moduli of \cite{Danielsson:2012et}. We conclude in Section \ref{concl} with a discussion of our results and some suggestions for future research. In Appendix \ref{app}, we review a no-go theorem against classical dS vacua proposed in \cite{Dasgupta:2014pma} and show that it can be evaded in the presence of O-planes.

\section{SU(3) Structure Orientifolds in Massive Type IIA}
\label{su3-IIA}

In this section, we consider SU$(3)$ structure compactifications of massive type IIA supergravity with O$6$-planes as well as $H$, $F_0$ and $F_2$ flux. The possibility of dS vacua in such models was analyzed in \cite{Caviezel:2008tf, Danielsson:2010bc, Danielsson:2011au} for various group and coset spaces, and a class of compactifications on orientifolds of the group space $\textrm{SU}(2) \times \textrm{SU}(2)$ was found to admit unstable dS critical points. In the following, we will first introduce the model and then show how the tachyon direction can be determined analytically. Finally, we will compare the tachyonic direction to the sgoldstino and give an explanation for its appearance in terms of a generalized version of the no-go theorem of \cite{Covi:2008ea}.

\subsection{The Model}

We consider the simple isotropic model of \cite{Danielsson:2010bc} with 4 intersecting O6-planes, $H$ and $F_2$ flux and a non-zero Romans mass $F_0$. Let us label the internal directions by the numbers $1,\ldots,6$. $H$ then has four components, which are along 456, 236, 134 and 125. The components of $F_2$ are along 16, 24 and 35, and the O$6$-planes are along 123, 145, 256 and 346.

The form fields are given by
\begin{align}
& \e^\phi F_0 = f_1, \quad \e^\phi F_2 = f_2 J, \quad H = f_5 \Omega_R + f_6 \hat W_3, \quad \e^\phi j = j_1\Omega_R + j_2\hat W_3,
\end{align}
where
\begin{align}
& J = a e^{16} - a e^{24} + a e^{35}, \\
& \Omega_R = v e^{456} + v e^{236} + \frac{a^6}{v^3} e^{134} + v e^{125}, \\
& \hat W_3 = -\frac{1}{2\sqrt{3}v^3} \left[ v^4 \left( e^{456} + e^{236} + e^{125} \right) -3a^6 e^{134} \right].
\end{align}
Furthermore, $j$ is the current 3-form coupling to the O6-plane sources and $a$, $v$, $f_1$, $f_2$, $f_5$, $f_6$, $j_1$ and $j_2$ are numbers determined by the equations of motion.
Here and in the following, we use the notation $e^{AB}=e^A \w e^B$, etc., where the basis 1-forms $e^A$ satisfy $\d e^A = \frac{1}{2}f^A\vphantom{}_{BC}\, e^B \w e^C$ with structure constants
\begin{equation}
f^1\vphantom{}_{23} = f^1\vphantom{}_{45} = f^2\vphantom{}_{56} = -f^3\vphantom{}_{46} = \frac{1}{2}, \qquad \textrm{cyclic}.
\end{equation}

As the orientifolding projects out off-diagonal deformations, the most general form of the internal metric is
\begin{equation}
g = \mathrm{diag} \left( g_{11}, g_{22}, g_{33},  g_{44}, g_{55}, g_{66} \right).
\end{equation}
The ansatz for the metric in Einstein frame is then
\begin{equation}
g = \e^{-\phi/2}a \, \mathrm{diag} \left( \frac{a^3}{v^2}, \frac{v^2}{a^3}, \frac{a^3}{v^2}, \frac{a^3}{v^2}, \frac{v^2}{a^3}, \frac{v^2}{a^3} \right). \label{IIA-metric}
\end{equation}
The internal curvature is defined as
\begin{equation}
R^{(6)} = \frac{1}{2} g^{EA} f^B\vphantom{}_{CE} f^C\vphantom{}_{AB} + \frac{1}{4} g^{LA} g^{BE} g_{DC} f^D\vphantom{}_{EL} f^C\vphantom{}_{AB}.
\end{equation}
We thus find
\begin{align}
R^{(6)} &= \frac{1}{2} \left({g^{11} + g^{22} + g^{33} + g^{44} + g^{55} + g^{66}}\right) \nl - \frac{1}{8} \left({ g^{11}g^{22}g_{33} + g^{11}g^{33}g_{22} + g^{22}g^{33}g_{11} + g^{11}g^{44}g_{55} + g^{11}g^{55}g_{44} + g^{44}g^{55}g_{11} }\right. \nl + \left.{g^{22}g^{55}g_{66} + g^{22}g^{66}g_{55} + g^{55}g^{66}g_{22} + g^{33}g^{44}g_{66} + g^{33}g^{66}g_{44} + g^{44}g^{66}g_{33}}\right) \nll
= - \frac{3}{8}\frac{\e^{\phi/2}v^6}{a^{10}} + \frac{3}{2}\frac{\e^{\phi/2}v^2}{a^4} +\frac{3}{8}\frac{\e^{\phi/2}a^2}{v^2}.
\end{align}

\subsection{dS Solutions}

The 10D equations of motion are
\begin{align}
& \e^{-\phi/2} R^{(4)} = -f_1^2-3f_2^2- 4 j_1, \\
& 0 = -\frac{3}{2}\frac{v^2}{a^4} -\frac{3}{8}\frac{a^2}{v^2} +\frac{3}{8}\frac{v^6}{a^{10}} + \frac{3}{4} \left( 4f_5^2+f_6^2\right) + \frac{3}{8} f_1^2 + \frac{15}{8} f_2^2 + \frac{9}{2} j_1, \\
& 0 = \frac{v^2}{a^4}\left(\frac{v^4}{4a^6} - \frac{1}{2}\right) + \frac{1}{2} \left( f_5-\frac{f_6}{2\sqrt{3}}\right)^2 + \frac{1}{2}\left( f_5+\frac{3f_6}{2\sqrt{3}}\right)^2 -\frac{1}{8} \left( 4f_5^2+f_6^2\right) + \frac{1}{16} f_1^2 \nl \quad + \frac{5}{16} f_2^2  + \frac{3}{4} j_1 + \frac{j_2}{2\sqrt{3}}, \\
& 0 = -\frac{1}{2} \left( 4f_5^2+f_6^2\right) + \frac{5}{4}f_1^2 + \frac{9}{4}f_2^2+3 j_1, \\
& 0 =\frac{1}{2} f_2a + f_1f_5v - f_1f_6\frac{v}{2\sqrt{3}}+ j_1v-  j_2\frac{v}{2\sqrt{3}}, \\
& 0 = \frac{3}{2}f_2a + f_1f_5\frac{a^6}{v^3} + f_1f_6\frac{3a^6}{2\sqrt{3}v^3}+  j_1\frac{a^6}{v^3}+  j_2\frac{3a^6}{2\sqrt{3}v^3}, \\
& 0 = -f_1f_2a^2 + \frac{1}{2}f_5 \frac{a^3}{v} - \frac{1}{2}f_6 \frac{a^3}{2\sqrt{3}v} + \frac{1}{2}f_5 \frac{v^3}{a^3} + \frac{1}{2}f_6 \frac{3v^3}{2\sqrt{3}a^3},
\end{align}
where $R^{(4)}$ denotes the external Ricci scalar in 10D Einstein frame. A one-parameter family of unstable dS critical points was numerically found in \cite{Danielsson:2010bc} in the range
\begin{equation}
v^4 = \frac{\sqrt{3} + w}{3\sqrt{3}-w}a^6, \qquad 4.553 < w < 3\sqrt{3}. \label{w-equation}
\end{equation}
At the two boundaries $w=3\sqrt{3}$ and $w\approx 4.553$, one obtains Minkowski vacua, while other values of $w$ yield AdS solutions.

An analytic expression for the family of dS solutions can be found as follows. For convenience, let us define the parameter
\begin{equation}
\epsilon = \frac{a^3}{v^2} \label{def-epsilon}
\end{equation}
in terms of which the scalar curvature of the internal space can be written as
\begin{equation}
R^{(6)} = - \frac{3\e^{\phi/2}}{8a\epsilon^3}  + \frac{3\e^{\phi/2}}{2a\epsilon} +\frac{3\e^{\phi/2}\epsilon}{8a}. \label{R6-epsilon}
\end{equation}
In order to obtain dS extrema in this model, $R^{(6)}$ must be negative and sufficiently small. This can be seen from the Einstein and dilaton equations, which imply
\begin{equation}
R^{(6)} = - \frac{9}{4}R^{(4)} -\frac{1}{2}\e^{3/2\phi}|F_2|^2,
\end{equation}
which is negative for $R^{(4)}> 0$. Comparing this to \eqref{R6-epsilon}, we conclude that $\epsilon$ is a small parameter for all dS critical points in this model. To make this more precise, we can substitute \eqref{def-epsilon} into \eqref{w-equation} to find the range
\begin{equation}
0 < \epsilon \lesssim 0.32. \label{ds-range}
\end{equation}
The existence of the small parameter $\epsilon$ is not a coincidence but could have been anticipated by noting that the internal space has topology $S^3\times S^3$. It can therefore only be negatively curved if it is significantly deformed away from the configuration of round spheres $g \sim \mathrm{diag} \left( 1,1,1,1,1,1 \right)$. Hence, for all dS critical points, there must be a large hierarchy between the different components of the internal metric, which is indeed verified by substituting \eqref{def-epsilon} into \eqref{IIA-metric},
\begin{equation}
g = \e^{-\phi/2} a \, \mathrm{diag} \left( \epsilon, \epsilon^{-1}, \epsilon, \epsilon, \epsilon^{-1}, \epsilon^{-1} \right).
\end{equation}

The above observations suggest that $\epsilon$ can be used as an expansion parameter. Solving the equations of motion in an expansion in $\epsilon$, one finds
\begin{align}
a &= \frac{3\sqrt{21}}{16\epsilon^2} - \frac{63}{64\epsilon} - \frac{367\sqrt{21}}{1792} - \frac{635\epsilon}{1024} + \mathcal{O}(\epsilon^2), \label{sol1} \\
f_2 &= \frac{2}{21^{1/4} \sqrt{3\epsilon}} + \frac{21^{3/4} \sqrt{\epsilon}}{12\sqrt{7}} - \frac{2585 \epsilon^{3/2}}{1344 \cdot 21^{1/4} \sqrt{3}} - \frac{32693 \epsilon^{5/2}}{4608 \cdot 21^{1/4} \sqrt{7}} + \mathcal{O}(\epsilon^{7/2}),\\
f_5 &= \frac{1}{4} - \frac{\sqrt{21}\epsilon}{7} - \frac{31\epsilon^2}{28} - \frac{29 \sqrt{21} \epsilon^3}{168} + \mathcal{O}(\epsilon^4),\\
f_6 &= \frac{\sqrt{3}}{2} + \frac{2\epsilon}{\sqrt{7}} - \frac{25\sqrt{3}\epsilon^2}{42} - \frac{47\epsilon^3}{12 \sqrt{7}} + \mathcal{O}(\epsilon^4), \label{sol3} \\
j_1 &= -\frac{1}{\sqrt{21} \epsilon} - \frac{1}{2} + \frac{67\epsilon}{42 \sqrt{21}} + \frac{61\epsilon^2}{63} + \mathcal{O}(\epsilon^3), \\
j_2 &= -\frac{2}{\sqrt{7}\epsilon} - \sqrt{3} - \frac{15\epsilon}{7\sqrt{7}} + \frac{62\sqrt{3}\epsilon^2}{63} + \mathcal{O}(\epsilon^3), \label{sol2}
\end{align}
where we have set $f_1=\e^\phi=1$ without loss of generality \cite{Danielsson:2010bc}.
In order to determine the exact convergence radius of the expansion, one would require the all-order analytic solution which is unfortunately not known. However, one can compare the expansion for $a,f_2,f_5,\ldots$ to a numerical solution for different values of $\epsilon$ and thus verify that the convergence radius of the expansion is $\approx 0.32$.
Hence, our expansion is valid for the whole range \eqref{ds-range} of dS critical points but ceases to converge as we approach the Minkowski point at $\epsilon\approx 0.32$.

\subsection{Scalar Potential}

In order to analyze the stability of the above dS critical points, let us now derive the 4D scalar potential of the compactified theory. We first introduce a dilaton modulus $\tau$ and two metric moduli $\sigma_1,\sigma_2$ probing the 134 and the 256 directions by redefining
\begin{gather}
\e^\phi \to \tau \e^\phi, \quad
g_{11} \to \sigma_1 g_{11}, \quad g_{22} \to \sigma_2 g_{22}, \quad g_{33} \to \sigma_1 g_{33}, \notag \\ g_{44} \to \sigma_1 g_{44}, \quad g_{55} \to \sigma_2 g_{55}, \quad g_{66} \to \sigma_2 g_{66}, \quad g_{\mu\nu} \to \frac{g_{\mu\nu}}{\text{vol}_6(\sigma_i)},
\end{gather}
where $\text{vol}_6(\sigma_i) = (\sigma_1\sigma_2)^{3/2} \int \d^{6}x \sqrt{g_{6}}$ and the last rescaling is required to stay in 4D Einstein frame. Note that our conventions are such that $\tau= \sigma_i = 1$ at a critical point. Hence, $g_{MN}$ and $\phi$ in the following expressions denote the on-shell values of the metric and the dilaton at a given solution. Also note that the internal Einstein equations only have two independent components in the isotropic model such that the two metric moduli we introduced capture all possible metric deformations preserving the isotropy property. There are also axionic moduli but they are not relevant for the tachyon.\footnote{Some of the solutions we discuss in this paper have one or more non-universal tachyons in addition to the ubiquitous one but we will not be concerned with them.} It was in fact already shown in \cite{Danielsson:2012et} that the tachyon in this model must lie in the moduli subspace spanned by the dilaton $\tau$, the 6D volume $(\sigma_1\sigma_2)^{3/2}$ and the O6-plane volume $\sigma_2^{3/2}$. Hence, it is sufficient for us to consider these 3 moduli.

The internal curvature with explicit moduli dependence thus reads
\begin{align}
R^{(6)}(\sigma_i) &= \frac{1}{2} \left({ \frac{g^{11}}{\sigma_1} + \frac{g^{22}}{\sigma_2} + \frac{g^{33}}{\sigma_1} + \frac{g^{44}}{\sigma_1}+ \frac{g^{55}}{\sigma_2}+ \frac{g^{66}}{\sigma_2} }\right) -
\frac{1}{8\sigma_2} \left(g^{11}g^{22}g_{33} + g^{11}g^{55}g_{44} \right. \nl + \left. g^{22}g^{55}g_{66} + g^{22}g^{33}g_{11} + g^{22}g^{66}g_{55} + g^{33}g^{66}g_{44} + g^{44}g^{55}g_{11} + g^{55}g^{66}g_{22} \right. \nl + \left. g^{44}g^{66}g_{33} \right) - 
\frac{\sigma_2}{8\sigma_1^2} \left(g^{11}g^{33}g_{22} + g^{11}g^{44}g_{55} + g^{33}g^{44}g_{66} \right).
\end{align}
Dimensionally reducing the 10D action, we find the 4D effective action
\begin{equation}
S \supset \int \d^{4}x \sqrt{- g_{4}} \left[{ R^{(4)} + \mathcal{L}_\text{kin}(\tau,\sigma_i) - V(\tau,\sigma_i) }\right]
\end{equation}
with the kinetic terms
\begin{equation}
\mathcal{L}_\text{kin}(\tau,\sigma_i) = - \frac{1
}{2} \frac{( \partial \tau)^2}{\tau^2} - \frac{15}{8} \frac{( \partial \sigma_1)^2}{\sigma_1^2} -\frac{15}{8}\frac{( \partial \sigma_2)^2}{\sigma_2^2}
- \frac{9}{4} \frac{(\partial \sigma_1)( \partial \sigma_2)}{\sigma_1\sigma_2}
\end{equation}
and the scalar potential
\begin{align}
V(\tau,\sigma_i) &= \frac{1}{\text{vol}_6(\sigma_i)} \left[{ V_R(\sigma_i) + \tau^{5/2} V_0 + \frac{\tau^{3/2}}{\sigma_1 \sigma_2} \, V_{2} + \frac{1}{\tau \sigma_1 \sigma_2^2} \, V_{31} + \frac{1}{\tau \sigma_1^3} \, V_{32} + \frac{1}{\tau \sigma_1 \sigma_2^2} \, V_{33} }\right. \nl + \left.{ \frac{1}{\tau \sigma_1 \sigma_2^2} \, V_{34}
+ \frac{\tau^{3/4}}{\sqrt{\sigma_1} \sigma_2 } \, V_{61}  + \frac{\tau^{3/4}}{\sqrt{\sigma_1} \sigma_2 } \, V_{62}  + \frac{\tau^{3/4}}{\sigma_1^{3/2} } \, V_{63}  + \frac{\tau^{3/4}}{\sqrt{\sigma_1} \sigma_2 } \, V_{64}}\right]. \label{scalar1}
\end{align}
Here, we have set $2\kappa_{10}^2=1$ for convenience and defined the coefficients
\begin{gather}
V_R(\sigma_i) = - R^{(6)}(\sigma_i), \quad V_0 = \frac{1}{2}\e^{5/2\phi} F_0^2, \quad V_{2} = \frac{1}{2}\e^{3/2\phi} | F_2|^2, \quad
V_{31} = \frac{1}{2}\e^{-\phi} |H^{(125)}|^2, \notag \\ V_{32} = \frac{1}{2}\e^{-\phi}| H^{(134)}|^2, \quad V_{33} = \frac{1}{2}\e^{-\phi}| H^{(236)}|^2, \quad V_{34} = \frac{1}{2}\e^{-\phi}| H^{(456)}|^2, \quad V_{61} =  \frac{\e^{3/4\phi} \mu^{(123)}_6}{\sqrt{g_{44}g_{55}g_{66}}}, \notag \\ V_{62} = \frac{\e^{3/4\phi} \mu^{(145)}_6}{\sqrt{g_{22}g_{33}g_{66}}}, \quad V_{63} = \frac{\e^{3/4\phi} \mu^{(256)}_6}{\sqrt{g_{11}g_{33}g_{44}}}, \quad V_{64} =\frac{\e^{3/4\phi} \mu^{(346)}_6}{\sqrt{g_{11}g_{22}g_{55}}},
\end{gather}
where $g_{MN}$ and $\phi$ again denote the on-shell values of the metric and the dilaton at the dS critical points. Our conventions are such that $\mu_6 > 0$ for net D-brane tension and $\mu_6 < 0$ for net O-plane tension. Note that the source terms do not contain any delta functions since we consider the smeared limit.
We thus find the on-shell expressions
\begin{gather}
V_R = \e^{\phi/2} \left( \frac{3}{8}\frac{v^6}{a^{10}} - \frac{3}{2}\frac{v^2}{a^4} - \frac{3}{8}\frac{a^2}{v^2} \right), \quad V_0 = \frac{1}{2} \e^{\phi/2} f_1^2, \quad V_{2} = \frac{3}{2}\e^{\phi/2} f_2^2, \notag \\ V_{31} = V_{33} = V_{34} = \frac{1}{2}\e^{\phi/2} \left( f_5-\frac{f_6}{2\sqrt{3}}\right)^2, \quad
V_{32} = \frac{1}{2}\e^{\phi/2} \left( f_5+\frac{3f_6}{2\sqrt{3}}\right)^2, \notag \\ V_{61} = V_{62} = V_{64} = \e^{\phi/2} \left( j_1-\frac{j_2}{2\sqrt{3}}\right), \quad V_{63} = \e^{\phi/2} \left( j_1+\frac{3j_2}{2\sqrt{3}}\right). \label{IIA-potential-terms}
\end{gather}

Let us finally state the K{\"{a}}hler potential and the superpotential of the compactified theory \cite{Caviezel:2008tf, Danielsson:2012et}. One finds
\begin{align}
& K = -\ln(z_1+\bar z_1)-3 \ln(z_2+\bar z_2)-3\ln(t+\bar t)+4\ln(2), \label{kaehler1} \\
& W = i \lambda_1 t^3 + 3t(\lambda_2t+z_1+z_2)-i \lambda_3 (z_1-3z_2) \label{super1}
\end{align}
in terms of the flux parameters $\lambda_i$ and the complex moduli $t$, $z_1$, $z_2$. Matching the F-term scalar potential obtained from \eqref{kaehler1} and \eqref{super1} to the scalar potential \eqref{scalar1}, one can relate the various parameters and moduli to those in our conventions. Putting again the axions on-shell, we find the following relations:
\begin{gather}
\lambda_1 = -\e^{-\phi}f_1, \quad \lambda_2 = \e^{-\phi}(4af_2- \mathrm{Im}(t) f_1), \quad \lambda_3 = -6f_5v+\sqrt{3}f_6v+2f_5\frac{a^6}{v^3}+\sqrt{3}f_6\frac{a^6}{v^3}, \notag \\
\mathrm{Re}(t)=4a \sqrt{\tau\sigma_1\sigma_2}, \quad \mathrm{Re}(z_1)=8\e^{-\phi} \frac{v^3}{a^3} \tau^{-1/4}\sigma_2^{3/2}, \quad \mathrm{Re}(z_2)=8\e^{-\phi} \frac{a^3}{v} \tau^{-1/4}\sigma_1 \sqrt{\sigma_2}, \notag \\ \mathrm{Im}(t) = -2f_5\frac{a^6}{v^3}-\sqrt{3}f_6\frac{a^6}{v^3}-2f_5v+\frac{1}{\sqrt{3}}f_6v, \notag \\ \mathrm{Im}(z_1) = \frac{\mathrm{Im}(t)}{4\lambda_3}\left(2\lambda_1 \mathrm{Im}(t)^2-3\lambda_2 \mathrm{Im}(t)+3\lambda_1\lambda_3 \mathrm{Im}(t)-6\lambda_2\lambda_3\right), \notag \\ \mathrm{Im}(z_2) = - \frac{\mathrm{Im}(t)}{4\lambda_3} \left(2\lambda_1 \mathrm{Im}(t)^2-3\lambda_2 \mathrm{Im}(t)-\lambda_1\lambda_3 \mathrm{Im}(t)+2\lambda_2\lambda_3\right). \label{relations1}
\end{gather}

\subsection{Tachyon}

We are now in a position to determine the tachyons. Our claim is that, for any $\epsilon$ in the range \eqref{ds-range}, there is an unstable direction $\alpha$, which lies in the subspace of the moduli space spanned by the dilaton and the 2 metric moduli. Let us formally define this direction as a simultaneous excitation of these moduli. We make the field redefinition $(\tau,\sigma_i)\to (\alpha,\beta_i)$ with
\begin{equation}
\tau = \alpha^t, \quad \sigma_1 = \alpha^{s_1} \beta_1, \quad \sigma_2 = \alpha^{s_2} \beta_2, \label{alpha}
\end{equation}
where $t,s_i$ are numbers whose values determine the direction of the $\alpha$ modulus inside the $(\tau,\sigma_i)$ hyperplane and which we leave undetermined for the moment. In the following, we will put the other moduli on-shell, $\beta_i=1$, and only consider the dependence of the scalar potential on the $\alpha$ modulus. It is convenient to parametrize the $\alpha$ direction by a vector
\begin{equation}
\left( t, s_1, s_2 \right) \label{vector-IIA}
\end{equation}
whose normalization is fixed by demanding that $\alpha$ is canonically normalized for small fluctuations around the critical point.\footnote{More precisely, our conventions are such that $\ln \alpha$ is canonically normalized. For small fluctuations, one then finds $\ln\alpha \approx \alpha - 1$ and $\mathcal{L}_\text{kin}=-\frac{1}{2}(\partial \ln \alpha)^2 \approx -\frac{1}{2}(\partial \alpha)^2$.} If we choose, e.g., $\left( 1, 0, 0 \right)$, $\alpha$ would coincide with the dilaton modulus, while the choice $\frac{1}{2\sqrt{3}} \left( 0, 1, 1 \right)$ would correspond to the overall volume modulus of the 6D internal space. For general $t$ and $s_i$, on the other hand, $\alpha$ is a mixed direction in the moduli space spanned by $\tau$ and $\sigma_i$.

Let us furthermore denote the one-by-one principal minor of the mass matrix along the direction $\alpha$ by
\begin{equation}
m_{\alpha}^2 = \left.{\frac{\partial^2 V(\alpha)}{\partial \alpha^2}}\right|_{\alpha=1}.
\end{equation}
We will now show that, for any dS critical point in the family of solutions labelled by $\epsilon$, there are directions for which $m_{\alpha}^2$ is manifestly negative. By Sylvester's criterion, this is sufficient to prove the existence of a tachyon. We could of course also compute the eigenvalues of the mass matrix to check for instabilities. However, diagonalizing the full mass matrix is often cumbersome, especially in models where the number of moduli is large, such that finding directions with negative principal minors and exploiting Sylvester's criterion can be very useful in practice. In many cases, the sgoldstino is such a useful direction \cite{Covi:2008ea} but in the context of classical dS vacua it is typically not. The main point of this section is to explicitly show that, for most $\epsilon$, the sgoldstino corresponds to none of the directions with negative principal minors (and, hence, cannot be used as a proxy to check for instabilities), and to explain why this is the case.

We first consider the limit $\epsilon\to 0$. Substituting the solution \eqref{sol1}--\eqref{sol2} into the scalar potential, we find
\begin{equation}
m_{\alpha}^2 = \frac{1024 (2s_1-4s_2+3t)^2}{11907} \epsilon^5 + \mathcal{O}(\epsilon^{6}).
\end{equation}
This is positive unless we set the leading order term to zero, $s_1 = 2s_2 - \frac{3}{2}t $. Substituting this back into $m_{\alpha}^2$, we find
\begin{equation}
m_{\alpha}^2 = \frac{1024 \sqrt{21} (t-6s_2)^2}{11907} \epsilon^6 + \mathcal{O}(\epsilon^7),
\end{equation}
which is again positive unless $t=6s_2$. With this choice, we arrive at
\begin{equation}
m_{\alpha}^2 = - \frac{2097152 \sqrt{21} s_2^2}{83349} \epsilon^8 + \mathcal{O}(\epsilon^{9}). \label{sgoldstino-mass}
\end{equation}
Hence, in the limit $\epsilon \to 0$, the mass matrix has a negative one-by-one minor along the direction
\begin{equation}
\frac{1}{8\sqrt{3}}\left(6,-7,1 \right), \label{LOsgoldstino}
\end{equation}
where we fixed the normalization of the vector by demanding that the $\alpha$ modulus is canonically normalized.

A closer look at the subleading corrections to the mass term \eqref{sgoldstino-mass} reveals that some of their coefficients are positive and quite large such that the expansion of the mass breaks down already for $\epsilon \ll 1$. This shows that the direction \eqref{LOsgoldstino} is not unstable for the whole family of dS solutions. One numerically verifies that it becomes stable at $\epsilon\approx 0.14$. Hence, the unstable direction is itself $\epsilon$-dependent and rotates away from \eqref{LOsgoldstino} for finite $\epsilon$. As one increases $\epsilon$, the leading order result \eqref{LOsgoldstino} ceases to be a good approximation since the expansion then converges more slowly to the exact direction.
However, we can easily also determine all possible directions with negative minors for any finite $\epsilon$.
To this end, we make a small detour and first determine all directions for which $m_{\alpha}^2=0$. From that, one can then deduce the existence of minors with $m_{\alpha}^2<0$, as we will see below.\footnote{Note that a vanishing minor does not imply a zero eigenvalue of the mass matrix. The fact that we find directions with $m_{\alpha}^2=0$ does therefore not mean that the solution contains massless modes.}

\begin{figure}[t]
\centering
\includegraphics[width=0.4\textwidth]{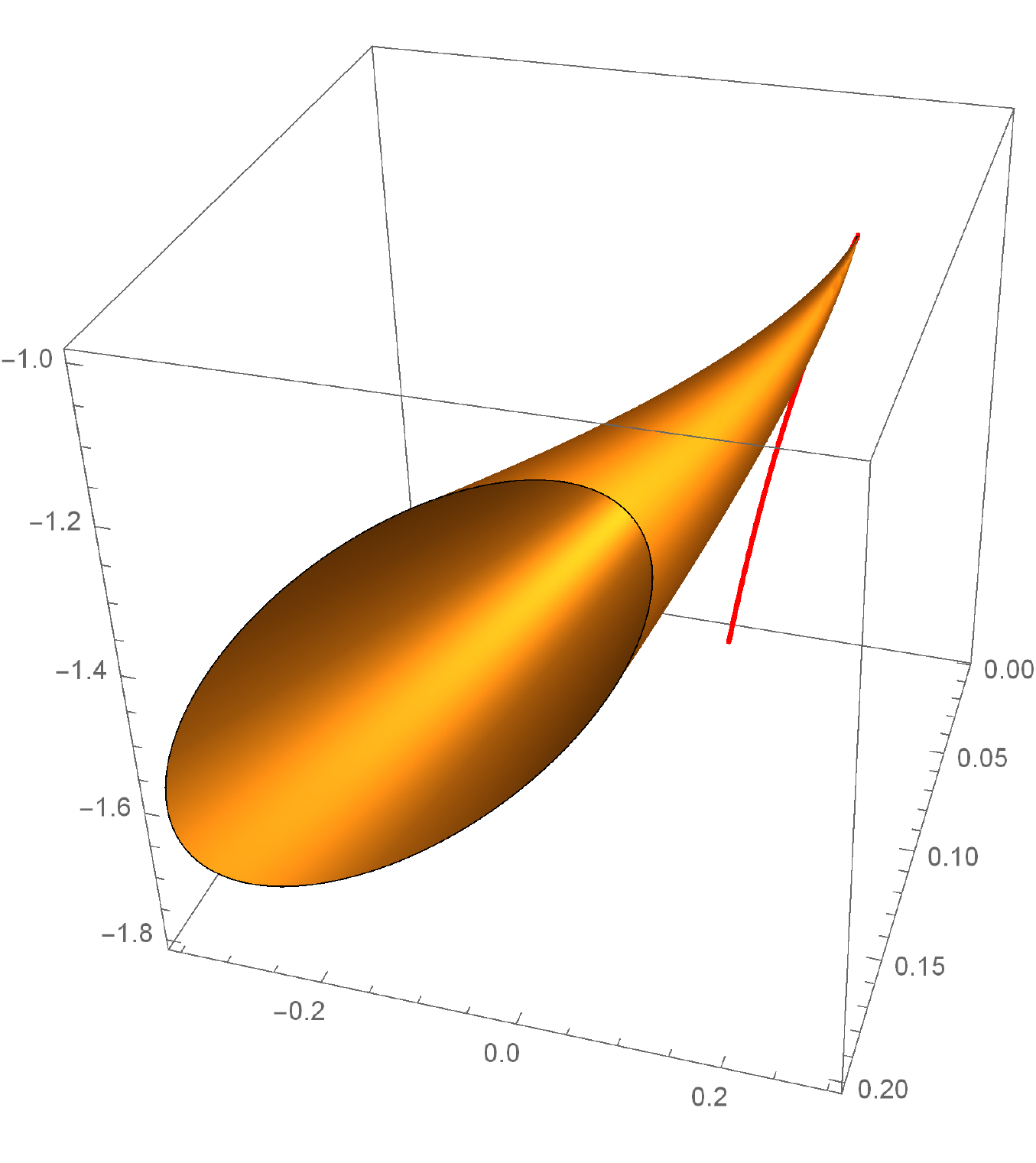}
\put(-110,5){$\frac{s_2}{t}$}
\put(-185,90){$\frac{s_1}{t}$}
\put(-5,45){$\epsilon$}
\caption{A plot of the $m_{\alpha}^2=0$ surface (yellow) and the sgoldstino (red) using the first few orders of the $\epsilon$-expansion. Directions which lie inside of the surface have negative principal minors and are therefore sufficient to prove the existence of a tachyon by Sylvester's criterion. For $\epsilon\to 0$, the surface shrinks to a point and aligns with the sgoldstino. \label{fig}}
\end{figure}

Upon using the equations of motion, we can write the minor along the $\alpha$ direction for general $t,s_1,s_2$ as
\begin{equation}
m_{\alpha}^2 = c_1 t^2 + c_2 s_1^2+c_3s_2^2+c_4ts_1+c_5ts_2+c_6s_1s_2 \label{m^2-general}
\end{equation}
with coefficients
\begin{align}
c_1 &= \frac{1}{a^3}\left(\frac{25}{4}V_0 + \frac{9}{4}V_{2} + 3 V_{31} + V_{32} + \frac{27}{16}V_{61} + \frac{9}{16}V_{63} \right), \label{coeff1} \\
c_2 &= \frac{1}{a^3}\left(\frac{9}{4}V_0 + \frac{25}{4}V_{2} + \frac{75}{4} V_{31} + \frac{81}{4}V_{32} + 12 V_{61} + 9 V_{63} + \frac{147}{32a\epsilon^3} - \frac{75}{8a\epsilon} - \frac{27\epsilon}{32a} \right), \\
c_3 &= \frac{1}{a^3}\left(\frac{9}{4}V_0 + \frac{25}{4}V_{2} + \frac{147}{4} V_{31} + \frac{9}{4}V_{32} + \frac{75}{4} V_{61} + \frac{9}{4} V_{63} + \frac{3}{32a\epsilon^3} - \frac{27}{8a\epsilon} - \frac{75\epsilon}{32a} \right), \\
c_4 &= \frac{1}{a^3}\left(- \frac{15}{2}V_0 - \frac{15}{2}V_{2} + 15 V_{31} + 9 V_{32} - 9 V_{61} -\frac{9}{2} V_{63} \right), \\
c_5 &= \frac{1}{a^3}\left(- \frac{15}{2}V_0 - \frac{15}{2}V_{2} + 21 V_{31} + 3 V_{32} -\frac{45}{4} V_{61} -\frac{9}{4} V_{63} \right), \\
c_6 &= \frac{1}{a^3}\left(\frac{9}{2}V_0 + \frac{25}{2}V_{2} + \frac{105}{2} V_{31} + \frac{27}{2}V_{32} + 30 V_{61} + 9 V_{63} + \frac{21}{16a\epsilon^3} - \frac{45}{4a\epsilon} - \frac{45\epsilon}{16a} \right). \label{coeff6}
\end{align}
Here, $t,s_1,s_2$ and $c_i,V_{ij},a$ should be read as being functions of $\epsilon$. Recall that the vector \eqref{vector-IIA} only has two independent components since its length is fixed by choosing a normalization for $\alpha$. In order to specify a direction inside the $(\tau,\sigma_i)$ space, it is therefore sufficient to determine, e.g., $\frac{s_1}{t}$ and $\frac{s_2}{t}$. Setting the left-hand side of \eqref{m^2-general} to zero and dividing by $t^2$, we get a quadratic equation for $\frac{s_1}{t}$ and $\frac{s_2}{t}$, with the solution
\begin{align}
& \frac{s_1}{t} = -\frac{c_6\frac{s_2}{t} + c_4 \pm \sqrt{\left(c_6\frac{s_2}{t}+c_4\right)^2-4c_2\left[c_1+c_5\frac{s_2}{t}+c_3(\frac{s_2}{t})^2\right]}}{2c_2}, \label{null-surface1} \\
& \left(c_6\frac{s_2}{t}+c_4\right)^2-4c_2\left[c_1+c_5\frac{s_2}{t}+c_3\left(\frac{s_2}{t}\right)^2\right] \ge 0, \label{null-surface2}
\end{align}
where the second condition follows from demanding that the square root in the first equation is real.
For each value of $\epsilon$, \eqref{null-surface1} and \eqref{null-surface2} parametrize a closed line in the $(\frac{s_1}{t},\frac{s_2}{t})$ space. Hence, the directions with $m_{\alpha}^2=0$ sweep out a surface in the space spanned by $(\epsilon,\frac{s_1}{t},\frac{s_2}{t})$ which has the useful property that $m_{\alpha}^2$ is positive for all $\frac{s_1}{t},\frac{s_2}{t}$ that lie outside of it, whereas it is negative for all $\frac{s_1}{t},\frac{s_2}{t}$ inside (cf. Fig. \ref{fig}).
At $\epsilon=0$, the surface shrinks to a point, namely \eqref{LOsgoldstino}.

In order to find a direction with a negative minor for a given $\epsilon$ in the range $0\le\epsilon\lesssim 0.32$, we select two points $(\frac{s_1}{t},\frac{s_2}{t})$ on the surface and then choose any point inbetween. A simple example is to take the two points for which the square root in \eqref{null-surface1} vanishes
and then take the middle between these two points,
\begin{equation}
\frac{s_1}{t} = -\frac{c_6\frac{s_2}{t} + c_4}{2c_2}, \quad \frac{s_2}{t} =  \frac{-c_4c_6 + 2c_2c_5}{c_6^2-4c_2c_3}. \label{tachyon-general}
\end{equation}

We can now substitute the definition of the coefficients \eqref{coeff1}--\eqref{coeff6} together with \eqref{IIA-potential-terms} and \eqref{sol1}--\eqref{sol2} and fix $t$ by demanding that $\alpha$ is canonically normalized. We thus find that the vector
\begin{align}
& \left( \frac{\sqrt{3}}{4} - \frac{169\sqrt{3}}{126}\epsilon^2-\frac{517\sqrt{7}}{441}\epsilon^3 + \frac{1081\sqrt{3}}{882}\epsilon^4, -\frac{7}{8\sqrt{3}} + \frac{101\sqrt{3}}{252}\epsilon^2+\frac{683\sqrt{7}}{882}\epsilon^3 + \frac{24193\sqrt{3}}{15876}\epsilon^4, \right. \notag \\ & \qquad\quad \left. \frac{1}{8\sqrt{3}} - \frac{53\sqrt{3}}{36}\epsilon^2-\frac{269\sqrt{7}}{126}\epsilon^3 + \frac{86351\sqrt{3}}{15876}\epsilon^4 \right)  + \mathcal{O}(\epsilon^5) \label{tachyon3}
\end{align}
is an example for an unstable direction in the $(\tau,\sigma_i)$ moduli space. As stated before, any other direction inside of the $m_\alpha^2=0$ surface would be suitable to detect the instability as well. By comparison with numerical solutions for different choices of $\epsilon$, one can verify that the above direction up to the order $\mathcal{O}(\epsilon^8)$ is indeed unstable until $\epsilon\approx 0.30$. This is ensured by the fact that its mass is given by \eqref{sgoldstino-mass}, up to small corrections. For even larger $\epsilon$, one would have to take into account higher orders in the expansion \eqref{tachyon3} since it then converges more slowly. We stress that it is in principle straightforward to do this and just a matter of computational effort.

Let us now compare the above to the sgoldstino $\phi$. The sgoldstino is the real part of the complex direction in moduli space along which supersymmetry is broken, i.e., for which $G_i = \partial_i \left( K + \ln |W|^2\right) \neq 0$.\footnote{A second sgoldstino is related to the imaginary part but not relevant for the tachyon here.}
Using \eqref{kaehler1}--\eqref{relations1} and \eqref{sol1}--\eqref{sol3}, we find that it is given by
\begin{align}
& \left( \frac{\sqrt{3}}{4} -\frac{\sqrt{3}}{12} \epsilon^2-\frac{95\sqrt{7}}{336} \epsilon^3-\frac{1783\sqrt{3}}{2016} \epsilon^4, -\frac{7}{8\sqrt{3}} -\frac{\sqrt{3}}{8} \epsilon^2 - \frac{29\sqrt{7}}{672} \epsilon^3 - \frac{61\sqrt{3}}{448} \epsilon^4, \right. \notag \\ & \qquad\quad \left. \frac{1}{8\sqrt{3}} + \frac{5\sqrt{3}}{24} \epsilon^2 - \frac{37\sqrt{7}}{672} \epsilon^3 - \frac{79\sqrt{3}}{1344} \epsilon^4 \right) + \mathcal{O}(\epsilon^5). \label{sgoldstino3}
\end{align}
Comparing this to \eqref{tachyon3}, we confirm that the tachyon indeed aligns with the sgoldstino in the singular limit $\epsilon \to 0$, as was already observed in \cite{Danielsson:2012et}. However, we also observe that the tachyon rotates away from the sgoldstino as $\epsilon$ is increased. Although the sgoldstino mass term starts out with a negative leading order contribution, it becomes stable already at small values of $\epsilon$ since its subleading coefficients are positive and rather large,
\begin{equation}
m_\phi^2 = - \frac{32768\sqrt{21}}{250047}\epsilon^8 + \frac{524288}{250047}\epsilon^9+ \frac{32620544\sqrt{21}}{5250987}\epsilon^{10}+\frac{991232}{7203}\epsilon^{11} + \mathcal{O}(\epsilon^{12}).
\end{equation}
In Fig. \ref{fig}, we have plotted the sgoldstino together with all directions for which the corresponding one-by-one minors in the mass matrix are negative. One indeed observes that the sgoldstino rotates away from the unstable region at larger $\epsilon$ and, hence, cannot be used to detect the instability anymore.

This can be explained as follows. First of all, let us check that, as claimed above, the dS solutions are close to a singular SUSY Minkowski point in moduli space. This can be seen from the contributions to the F-term scalar potential, whose leading order behavior is
\begin{equation}
\e^K g^{i \bar{\jmath}} D_iW D_{\bar{\jmath}} \overline W \sim \epsilon^5, \qquad -3 \e^K |W|^2 \sim \epsilon^5.
\end{equation}
Hence, for $\epsilon \to 0$, the SUSY equations are satisfied and the scalar potential vanishes. We therefore expect that the tachyon should align with the sgoldstino in this singular limit, and indeed this is the case.

But why is the tachyon not completely aligned with the sgoldstino at finite $\epsilon$? The reason for this is simple. The mass matrix in a general model with broken SUSY takes the schematic form\footnote{Here, we are again only concerned with the real part of the moduli space.}
\begin{equation} \label{sgoldstino-mixing}
\begin{pmatrix}
   m_\phi^2 & \mu^2 \\
   \mu^2 & M^2
\end{pmatrix},
\end{equation}
where $m_\phi$ is the sgoldstino mass, $\mu$ denotes possible off-diagonal terms and $M$ stands for the masses of an arbitrary number of other moduli, which we assume to be large. The eigenvalues are then
\begin{equation}
m_\phi^2-\frac{\mu^4}{M^2} + \mathcal{O}\left(M^{-4}\right), \qquad M^2+\frac{\mu^4}{M^2} + \mathcal{O}\left(M^{-4}\right)
\end{equation}
with eigenvectors
\begin{equation}
\left(1,-\frac{\mu^2}{M^2}\right) + \mathcal{O}\left(M^{-4}\right), \qquad \left(\frac{\mu^2}{M^2},1\right) + \mathcal{O}\left(M^{-4}\right).
\end{equation}
Assuming that the SUSY breaking scale is small compared to the moduli masses, $m_\phi^2 \ll M^2$, and that the mixing between the sgoldstino and the orthogonal moduli can be tuned to vanish, $\mu \to 0$, the eigenvector corresponding to the smallest eigenvalue always points into the sgoldstino direction and, hence, $m_\phi^2>0$ is sufficient to guarantee a stable dS vacuum \cite{Covi:2008ea}. However, in a general mass matrix with $\mu \neq 0$, the smallest eigenvalue receives corrections from off-diagonal terms and the corresponding eigenvector rotates away from the sgoldstino. This is particularly relevant in classes of string compactifications where the scalar potential is not sufficiently generic to allow these off-diagonal terms to be tuned small.

In a field basis in which the sgoldstino and the two orthogonal fields are canonically normalized, the mass matrix of our model takes the form
\begin{equation}
V_{ij} \sim \begin{pmatrix}
   \mathcal{O}(\epsilon^8) & \mathcal{O}(\epsilon^7) & \mathcal{O}(\epsilon^7) \\
   \mathcal{O}(\epsilon^7) & \mathcal{O}(\epsilon^5) & \mathcal{O}(\epsilon^5) \\
   \mathcal{O}(\epsilon^7) & \mathcal{O}(\epsilon^5) & \mathcal{O}(\epsilon^5)
\end{pmatrix},
\end{equation}
where the first row contains the sgoldstino mass term $m_\phi^2$ and the off-diagonal components $V_{\phi j}$.
Hence, there is indeed a large mixing between the sgoldstino and the other directions in field space.
As $\epsilon$ is increased, the tachyon eigenvector then rotates away from the sgoldstino and the corresponding eigenvalue is negative even though the principal minor along the sgoldstino direction is positive. A proper generalization of the sgoldstino no-go theorem of \cite{Covi:2008ea} to the case of a general mass matrix thus predicts that a tachyon, if present, should be close to, but not completely aligned with the sgoldstino, where the separation between the two directions is expected to be of the order $\mu^2/M^2 \sim \mathcal{O}(\epsilon^2)$. This is exactly reproduced by our above result.

\section{SU(2) Structure Orientifolds in Type IIB}
\label{su2-IIB}

In this section, we will show that our explanation for the tachyons also holds for an SU$(2)$ structure compactification of type IIB supergravity with O5/O7-planes and RR fluxes. In \cite{Caviezel:2009tu}, several compactifications of this type on group and coset spaces were analyzed and scanned for the existence of dS vacua. The authors then found that a compactification on an orientifold of the group space $\textrm{SU}(2) \times \textrm{SU}(2)$ indeed admits a dS critical point. The solution is unstable, and the reason for this instability has previously not been understood. One should stress that this model is not related by T-duality to a geometric type IIA compactification such as the one discussed in the previous section \cite{Caviezel:2009tu}. The fact that our argument also explains the tachyon here is therefore a non-trivial check which makes us confident that we have found a model-independent explanation for the instabilities.

\subsection{The Model}

The model we consider contains O5-planes and O7-planes as well as non-trivial RR and NSNS field strengths $F_1$, $F_3$ and $H$. Let us again label the internal directions by the numbers $1,\ldots,6$. The components of $F_1$ are then along the 1 and 2 directions, the components of $F_3$ are along 136, 236, 145 and 245, the components of $H$ are along 134, 234, 156, 256, the O$5$-planes are along 34 and 56 and the O$7$-planes are along 1235 and 1246.

The ansatz for the form fields is\footnote{Our definition of the flux parameters differs from the one in \cite{Caviezel:2009tu} because of different sign conventions and the fact that we work with improved field strengths $F=\d C - H\w C$.}
\begin{align}
F_1 &= m_1 e^1 + m_2e^2, \\
F_3 &= f_{1} (e^{136}+e^{245})+ f_{2} (e^{145}+e^{236}) + c_{1} (e^{145}-e^{236}) + c_{2} (e^{136}-e^{245}), \\
H &= b_1 (-e^{156}+e^{234})+b_2 (-e^{134}+e^{256}),
\end{align}
where the 1-forms $e^A$ satisfy $\d e^A = \frac{1}{2}f^A\vphantom{}_{BC}\, e^B \w e^C$ with structure constants
\begin{equation}
f^1\vphantom{}_{35} = f^2\vphantom{}_{46} = 1, \qquad \textrm{cyclic}. \label{structure-IIB}
\end{equation}

For $g_{mn} = \delta_{mn}$, the internal space would be a direct product of two round 3-spheres. However, in order to obtain dS vacua, an internal space with negative curvature is required. Hence, at any dS critical point, the internal space must be deformed away from a product of round spheres. The deformations can in general be both rescalings of the diagonal entries away from 1 and non-zero off-diagonal modes. For the orientifold in this particular model, the most general form of the internal metric reads \cite{Petrini:2013ika}
\begin{equation}
g = \begin{pmatrix}
   g_{11} & g_{12} & 0 & 0 & 0 & 0 \\
   g_{12} & g_{22} & 0 & 0 & 0 & 0 \\
   0 & 0 & g_{33} & 0 & 0 & 0 \\
   0 & 0 & 0 & g_{44} & 0 & 0 \\
   0 & 0 & 0 & 0 & g_{55} & 0 \\
   0 & 0 & 0 & 0 & 0 & g_{66}
\end{pmatrix}.
\end{equation}
For later convenience, let us express the metric components in terms of the variables used in \cite{Caviezel:2009tu}. Using the definition of \cite{Petrini:2013ika} of the globally defined forms on SU(2) structure manifolds together with the expressions of \cite{Caviezel:2009tu}, we find the Einstein frame metric
\begin{equation}
g = \e^{-\phi/2} \begin{pmatrix}
  \scriptstyle{L^2} & \scriptstyle{-L^2y} & 0 & 0 & 0 & 0 \\
   \scriptstyle{-L^2y} & \scriptstyle{L^2(x^2+y^2)} & 0 & 0 & 0 & 0 \\
   0 & 0 & \frac{k_1^2 k_2 L^2 x}{\e^{2\phi}u_2v_2} & 0 & 0 & 0 \\
   0 & 0 & 0 & \frac{k_2v_2}{u_2L^2x} & 0 & 0 \\
   0 & 0 & 0 & 0 & \frac{k_2u_2L^2x}{v_2} & 0 \\
   0 & 0 & 0 & 0 & 0 & \frac{\e^{2\phi}u_2v_2}{k_2L^2x} 
\end{pmatrix} \label{metric2}
\end{equation}
with
\begin{equation}
L^2=\frac{1}{x} \sqrt{-\frac{v_1v_2}{u_1u_2}}, \qquad \e^{2\phi} = \sqrt{-\frac{L^4x^2k_1^2k_2^2}{u_1u_2v_1v_2}}.
\end{equation}

Finally, the components of the Ricci tensor can be computed using the formula
\begin{equation}
R_{AD} = \frac{1}{4} g^{CE} g^{BF} g_{AG} g_{DH} f^G\vphantom{}_{FE} f^H\vphantom{}_{BC} - \frac{1}{2} g_{BE} g^{CF} f^B\vphantom{}_{AC} f^E\vphantom{}_{DF} - \frac{1}{2} f^C\vphantom{}_{AB} f^B\vphantom{}_{DC},
\end{equation}
which yields
\begin{align}
R_{11} &= \frac{1}{2} \left( g^{55}g^{33}g_{11}g_{11} + g^{66}g^{44}g_{12}g_{12} - g_{33}g^{55} - g_{55}g^{33} \right) + 1, \\
R_{12} &= \frac{1}{2} \left( g^{55}g^{33}g_{11}g_{21} + g^{66}g^{44}g_{12}g_{22} \right), \\
R_{22} &= \frac{1}{2} \left( g^{66}g^{44}g_{22}g_{22} + g^{55}g^{33}g_{21}g_{21} - g_{44}g^{66} - g_{66}g^{44} \right) + 1, \\
R_{33} &= \frac{1}{2} \left( g^{55}g^{11}g_{33}g_{33} - g_{11}g^{55} - g_{55}g^{11} \right) + 1, \\
R_{44} &= \frac{1}{2} \left( g^{66}g^{22}g_{44}g_{44} - g_{22}g^{66} - g_{66}g^{22} \right) + 1, \\
R_{55} &= \frac{1}{2} \left( g^{33}g^{11}g_{55}g_{55} - g_{11}g^{33} - g_{33}g^{11} \right) + 1, \\
R_{66} &= \frac{1}{2} \left( g^{44}g^{22}g_{66}g_{66} - g_{22}g^{44} - g_{44}g^{22} \right) + 1.
\end{align}

\subsection{dS Solution}

The 10D equations of motion are
\begin{align}
&R^{(4)} = -\frac{1}{2}\e^{-\phi}|H|^2-\frac{1}{2}\e^{\phi}|F_3|^2 - \frac{1}{2} \e^{\phi/2}\left(  \frac{\mu_5^{(34)}}{\sqrt{(\det g_{ab}) g_{55}g_{66}}}+\frac{\mu_5^{(56)}}{\sqrt{(\det g_{ab}) g_{33}g_{44}}}\right), \\
&0 = -\frac{1}{2}\e^{-\phi}|H|^2+\e^{2\phi}|F_1|^2+\frac{1}{2}\e^{\phi}|F_3|^2 + \frac{1}{2} \e^{\phi/2}\left(  \frac{\mu_5^{(34)}}{\sqrt{(\det g_{ab}) g_{55}g_{66}}}+\frac{\mu_5^{(56)}}{\sqrt{(\det g_{ab}) g_{33}g_{44}}}\right) \notag \\ & \qquad\quad + \e^\phi\left(  \frac{\mu_7^{(1235)}}{\sqrt{g_{44}g_{66}}} +\frac{\mu_7^{(1246)}}{\sqrt{g_{33}g_{55}}}\right), \\
&R_{mn} = \frac{1}{2}\e^{-\phi}|H|^2_{mn} + \frac{1}{2}\e^{2\phi}|F_1|^2_{mn} + \frac{1}{2}\e^{\phi}|F_3|^2_{mn} -\frac{1}{8}g_{mn}\e^{-\phi}|H|^2 -\frac{1}{8}g_{mn}\e^{\phi}|F_3|^2 \notag \\ & \qquad\quad  + \frac{3}{8}g_{mn} \e^{\phi/2}\left(  \frac{\mu_5^{(34)}}{\sqrt{(\det g_{ab}) g_{55}g_{66}}}+\frac{\mu_5^{(56)}}{\sqrt{(\det g_{ab}) g_{33}g_{44}}}\right) \notag \\ & \qquad\quad +\frac{1}{2}g_{mn}\e^\phi\left(  \frac{\mu_7^{(1235)}}{\sqrt{g_{44}g_{66}}} +\frac{\mu_7^{(1246)}}{\sqrt{g_{33}g_{55}}}\right) + \frac{1}{2}t_{mn}, \\
&\d F_1 =  \mu_7^{(1235)} e^{46} - \mu_7^{(1246)} e^{35}, \\
&\d F_3 = H \w F_1 - \mu_5^{(34)} e^{1256} - \mu_5^{(56)} e^{1234}, \\
&\d \left(\e^{-\phi}\star_6 H \right) + \e^\phi \star_6 F_3 \w F_1 = 0, \\
&\d \left(\e^{\phi}\star_6 F_3 \right) = 0,
\end{align}
where $a,b=1,2$ and
\begin{align}
& t_{ab} = -\e^\phi \left( \frac{\mu_7^{(1235)}}{\sqrt{g_{44}g_{66}}}+\frac{\mu_7^{(1246)}}{\sqrt{g_{33}g_{55}}}\right)g_{ab}, \notag \\
& t_{33} = -\left( \e^\phi \frac{\mu_7^{(1235)}}{\sqrt{g_{44}g_{66}}}+\e^{\phi/2}\frac{\mu_5^{(34)}}{\sqrt{(\det g_{ab}) g_{55}g_{66}}} \right)g_{33}, \notag \\
& t_{44} = -\left( \e^\phi \frac{\mu_7^{(1246)}}{\sqrt{g_{33}g_{55}}}+\e^{\phi/2}\frac{\mu_5^{(34)}}{\sqrt{(\det g_{ab}) g_{55}g_{66}}} \right)g_{44}, \notag \\
& t_{55} = -\left( \e^\phi \frac{\mu_7^{(1235)}}{\sqrt{g_{44}g_{66}}}+\e^{\phi/2}\frac{\mu_5^{(56)}}{\sqrt{(\det g_{ab}) g_{33}g_{44}}} \right)g_{55}, \notag \\
& t_{66} = -\left( \e^\phi \frac{\mu_7^{(1246)}}{\sqrt{g_{33}g_{55}}}+\e^{\phi/2}\frac{\mu_5^{(56)}}{\sqrt{(\det g_{ab}) g_{33}g_{44}}} \right)g_{66}.
\end{align}

The numerical dS solution is \cite{Caviezel:2009tu}
\begin{gather}
x=0.267585, \quad k_1=1.76189, \quad k_2=1.97367, \quad u_1=2.38469, \notag \\ u_2=0.0406036, \quad v_1=-0.00820371, \quad v_2=0.0512969, \quad
y=0.624470, \notag \\ m_1=-1.26529, \quad m_2=1.92725, \quad b_1=6.22664,\quad b_2=3.41528, \notag \\ c_1=-3.16142, \quad c_2=-0.691846,\quad f_1=5.44653, \quad f_2=-6.70494. \label{numerical-dS}
\end{gather}
Plugging this into \eqref{metric2}, we find
\begin{equation}
g = \begin{pmatrix}
  \scriptstyle{0.100641} & \scriptstyle{-0.0628475} & 0 & 0 & 0 & 0 \\
   \scriptstyle{-0.0628475} & \scriptstyle{0.0464525} & 0 & 0 & 0 & 0 \\
   0 & 0 & \scriptstyle{2.20576} & 0 & 0 & 0 \\
   0 & 0 & 0 & \scriptstyle{15.4502} & 0 & 0 \\
   0 & 0 & 0 & 0 & \scriptstyle{0.0420713} & 0 \\
   0 & 0 & 0 & 0 & 0 & \scriptstyle{0.234840}
\end{pmatrix}. \label{metric3}
\end{equation}
We checked that the above solution satisfies the 10D equations of motion, thus confirming that the 4D solution of \cite{Caviezel:2009tu} is a consistent truncation as expected. It should be noted that the string frame volume and the dilaton take the values $\int \d^6 x \sqrt{g_6}\,\e^{3\phi/2} = 0.229249$ and $\e^\phi = 5.99279$ in this solution such that we are not in the large volume/weak coupling regime \cite{Caviezel:2009tu}. The solution should therefore be viewed as a toy example, which is useful to study the general stability issues at dS critical points of type IIB supergravity.\footnote{Due to the scale invariance of the supergravity equations, it is often possible to rescale a given solution such that one obtains a large volume/weak coupling solution from a small volume/strong coupling one. However, this procedure typically involves a rescaling of the tension and the charges of the localized sources, which is problematic in the presence of O-planes if flux/charge quantization is imposed. Since O-planes cannot be stacked, increasing their number is not possible without changing the internal geometry.}

One immediately observes in \eqref{metric3} that, as in the type IIA examples discussed above, there is a large hierarchy between the different metric components such that we are far away from the configuration of round spheres. These metric deformations tend to make the internal curvature negative, which is required in order to get dS vacua. However, they also move the solution close to a singular point in moduli space where some of the cycle volumes blow up or shrink to zero size and the scale of SUSY breaking is small compared to (some of) the moduli masses. This suggests that our argument for the appearance of the tachyons should again hold in this model, which we will indeed confirm below. Let us finally note that it should in principle be possible in this model to compute an analytic (family of) dS solution(s) order by order in a small SUSY-breaking parameter $\epsilon$, analogous to what we did in Section \ref{su3-IIA}. Unfortunately, we did not succeed in doing so since the computation becomes very expensive already at the first few orders due to the much more complicated equations of motion. We therefore use the numerical solution of \cite{Caviezel:2009tu} in the following, which is sufficient for our purpose.

\subsection{Scalar Potential}

Let us now again derive the 4D scalar potential. As in the previous section, we first introduce a dilaton modulus $\tau$ and metric moduli $\sigma_i$ by redefining
\begin{gather}
\e^\phi \to \tau \e^\phi, \quad
g_{ab} \to \sigma_1 g_{ab}, \quad g_{33} \to \sigma_3 g_{33}, \quad g_{44} \to \sigma_4 g_{44}, \notag \\ g_{55} \to \sigma_5 g_{55}, \quad g_{66} \to \sigma_6 g_{66}, \quad g_{\mu\nu} \to \frac{g_{\mu\nu}}{\text{vol}_6(\sigma_i)},
\end{gather}
where $a,b=1,2$ and $\text{vol}_6(\sigma_i) = \sigma_1 \sqrt{\sigma_3 \sigma_4 \sigma_5 \sigma_6} \int \d^{6}x \sqrt{g_{6}}$. Note that, for simplicity, we do not consider all possible deformations of the 2D subspace along 12 where the metric is not diagonal. Instead, we only introduced an overall rescaling mode $\sigma_1$ for these metric components. One can check that the real component of the sgoldstino points almost exclusively into the $(\tau,\sigma_i)$ subspace of the full moduli space such that these moduli are sufficient for our argument. The rescaling of the external metric $g_{\mu\nu}$ is again required in order to end up in 4D Einstein frame. Furthermore, our conventions are again such that $\tau= \sigma_i = 1$ at the critical point and, hence, $g_{MN}$ and $\phi$ in the following expressions denote the on-shell values of the metric and the dilaton. Finally, we will not explicitly consider the dependence of the scalar potential on the axions since those are not relevant for the tachyon and can be put on-shell.

Dimensionally reducing the 10D action and using the above definitions, we obtain the 4D effective action
\begin{equation}
S \supset \int \d^{4}x \sqrt{- g_{4}} \left[{ R^{(4)} + \mathcal{L}_\text{kin}(\tau,\sigma_i) - V(\tau,\sigma_i) }\right]
\end{equation}
with the kinetic terms
\begin{align} \label{kin-IIB}
\mathcal{L}_\text{kin}(\tau,\sigma_i) &= - \frac{1}{2}\frac{(\partial \tau)^2}{\tau^2} - \frac{(\partial \sigma_1)^2}{\sigma_1^2} -\frac{3}{8}\frac{(\partial \sigma_3)^2}{\sigma_3^2} -\frac{3}{8}\frac{(\partial \sigma_4)^2}{\sigma_4^2}-\frac{3}{8}\frac{(\partial \sigma_5)^2}{\sigma_5^2}-\frac{3}{8}\frac{(\partial \sigma_6)^2}{\sigma_6^2} \nl
- \frac{1}{2} \frac{(\partial \sigma_1)(\partial \sigma_3)}{\sigma_1\sigma_3}
- \frac{1}{2} \frac{(\partial \sigma_1)(\partial \sigma_4)}{\sigma_1\sigma_4}
- \frac{1}{2} \frac{(\partial \sigma_1)(\partial \sigma_5)}{\sigma_1\sigma_5}
- \frac{1}{2} \frac{(\partial \sigma_1)(\partial \sigma_6)}{\sigma_1\sigma_6}\nl
- \frac{1}{4} \frac{(\partial \sigma_3)(\partial \sigma_4)}{\sigma_3\sigma_4}
- \frac{1}{4} \frac{(\partial \sigma_3)(\partial \sigma_5)}{\sigma_3\sigma_5}
- \frac{1}{4} \frac{(\partial \sigma_3)(\partial \sigma_6)}{\sigma_3\sigma_6}
- \frac{1}{4} \frac{(\partial \sigma_4)(\partial \sigma_5)}{\sigma_4\sigma_5}\nl
- \frac{1}{4} \frac{(\partial \sigma_4)(\partial \sigma_6)}{\sigma_4\sigma_6}
- \frac{1}{4} \frac{(\partial \sigma_5)(\partial \sigma_6)}{\sigma_5\sigma_6}
\end{align}
and the scalar potential
\begin{align}
\label{V2-typeIIB}
V(\tau,\sigma_i) &= \frac{1}{\text{vol}_6(\sigma_i)} \left[{ V_R(\sigma_i) + \frac{\tau}{\sigma_1 \sigma_3\sigma_6} \, V_{31} + \frac{\tau}{\sigma_1 \sigma_4\sigma_5} \, V_{32} + \frac{1}{\tau \sigma_1 \sigma_3\sigma_4} \, V_{33} + \frac{1}{\tau \sigma_1 \sigma_5\sigma_6} \, V_{34} }\right. \nl \left.{ + \frac{\tau^2}{\sigma_1} V_1 + \frac{\sqrt{\tau}}{\sigma_1 \sqrt{ \sigma_5 \sigma_6} } \, V_{51} + \frac{\sqrt{\tau}}{\sigma_1 \sqrt{\sigma_3 \sigma_4}} \, V_{52} + \frac{\tau}{\sqrt{\sigma_4 \sigma_6}} \, V_{71} + \frac{\tau}{\sqrt{\sigma_3 \sigma_5}} \, V_{72} }\right],
\end{align}
where we set $2\kappa_{10}^2=1$ for convenience and defined
\begin{gather} 
V_R(\sigma_i) = - R^{(6)}(\sigma_i), \quad V_{31} = \frac{1}{2} \e^{\phi} | F^{(136)}_3 + F^{(236)}_3|^2, \quad V_{32} = \frac{1}{2}\e^{\phi} | F^{(145)}_3 + F^{(245)}_3|^2, \notag \\ V_{33} = \frac{1}{2}\e^{-\phi} | H^{(134)} + H^{(234)}|^2, \quad V_{34} = \frac{1}{2}\e^{-\phi} | H^{(156)} + H^{(256)}|^2, \quad V_1 = \frac{1}{2} \e^{2\phi} |F_1|^2, \notag \\ V_{51} = \frac{\e^{\phi/2} \mu^{(34)}_5}{\sqrt{(\det g_{ab})g_{55}g_{66}}}, \quad V_{52} = \frac{\e^{\phi/2} \mu^{(56)}_5}{\sqrt{(\det g_{ab})g_{33}g_{44}}}, \quad V_{71} = \frac{\e^{\phi} \mu^{(1235)}_7}{\sqrt{g_{44}g_{66}}}, \quad V_{72} = \frac{\e^{\phi} \mu^{(1246)}_7}{\sqrt{g_{33}g_{55}}}. \label{V-typeIIB}
\end{gather}
Our conventions are such that $\mu_5 > 0,\mu_7 > 0$ for net D-brane tension and $\mu_5 < 0,\mu_7 < 0$ for net O-plane tension.

The K{\"{a}}hler potential and the superpotential are \cite{Caviezel:2009tu}
\begin{align}
K &= -\ln \left[ (T+\bar T)(t_1+\bar t_1)(t_2+\bar t_2)\right] \nl - \ln\left[ -\frac{1}{16} (z_1+\bar z_1)(z_2+\bar z_2)(w_1+\bar w_1)(w_2+\bar w_2) \right], \label{kaehler-IIB} \\
W &= -\frac{i}{2}\left[ f_{31}(-t_1+it_2T)+f_{32}(-t_2+it_1T)-im_2t_1t_2-m_1t_1t_2T+it_1z_2+it_2z_1 \right. \nl \left. -t_1z_1T-t_2z_2T-iw_1T-w_2\right], \label{super-IIB}
\end{align}
where the complex moduli are defined as
\begin{gather}
T=x+iy, \quad t_1=k_1+ib_1, \quad t_2=k_2+ib_2, \quad w_1=v_1+ic_{41}, \quad w_2=v_2+ic_{42}, \notag \\ z_1=u_1+ic_{21}, \quad z_2=u_2+ic_{22}
\end{gather}
and the axions and flux parameters are related to our parameters as follows:
\begin{gather}
c_{41} = m_1b_1b_2-b_1c_1-b_2c_2-b_2f_1-b_1f_2, \quad c_{42} = -m_2b_1b_2-b_2c_1-b_1c_2+b_1f_1+b_2f_2, \notag \\
c_{21}=c_1+\frac{1}{2}(m_2b_1-m_1b_2), \quad c_{22}=c_2-\frac{1}{2}(m_1b_1-m_2b_2), \notag \\
f_{31} = -f_1+\frac{1}{2}(m_1b_1+m_2b_2), \quad f_{32} = -f_2+\frac{1}{2}(m_2b_1+m_1b_2).
\end{gather}
One can check that, using these relations, the F-term scalar potential agrees with the scalar potential obtained from the dimensional reduction.

\subsection{Tachyon}

We will now show that the dS solution \eqref{numerical-dS} has a tachyon $\alpha$ in the $(\tau,\sigma_i)$ space which is close to, but not completely aligned with the sgoldstino. The reason for this behavior is the same as in the type IIA model: due to the large hierarchy in the metric \eqref{metric3}, the dS critical point is close to a singular SUSY point in moduli space such that the sgoldstino and nearby directions are the most dangerous region in the moduli space. Mixing effects in the mass matrix between the sgoldstino and the orthogonal modes then cause the tachyon to slightly rotate away from the sgoldstino.

We define $\alpha$ as a simultaneous excitation of the moduli,
\begin{equation}
\tau = \alpha^t, \quad \sigma_i = \alpha^{s_i}\beta_i, \label{alpha-typeIIB}
\end{equation}
where $t,s_i$ are numbers whose values determine the direction of the $\alpha$ modulus inside the $(\tau,\sigma_i)$ hyperplane.
We again parametrize this direction by a vector
\begin{equation}
\left( t, s_1, s_3, s_4, s_5, s_6 \right),
\end{equation}
whose normalization is fixed by demanding that $\alpha$ is canonically normalized for small fluctuations around the critical point.

One verifies that the contributions to the F-term scalar potential are
\begin{equation}
\e^K g^{i \bar{\jmath}} D_iW D_{\bar{\jmath}} \overline W = 34338.8, \qquad -3 \e^K |W|^2 = -34310.7.
\end{equation}
Since neither the individual terms nor their sum is small in absolute numbers, one might wonder in which sense the dS critical point is close to a SUSY Minkowski point. However, we will see below that the claim is true in the sense that the sgoldstino mass is small compared to the masses of (some of) the orthogonal moduli.

Using \eqref{kaehler-IIB} and \eqref{super-IIB}, we find that the sgoldstino direction is
\begin{gather}
\left( 0.27, -0.43, 0.15, 0.14, -0.44, -0.43 \right)
\end{gather}
with
\begin{equation}
m_\phi^2 = 7.6 \cdot 10^2,
\end{equation}
where we have rounded to 2 significant digits. Hence, the sgoldstino itself is stable. The tachyon, on the other hand, is given by
\begin{equation}
\left(0.12, -0.62, -0.0062, 0.11, -0.24, -0.12 \right) \label{IIB-tachyon}
\end{equation}
with
\begin{equation}
m_\alpha^2 = -1.3 \cdot 10^2.
\end{equation}
At first sight, it does not look as if the tachyon is closely related to the sgoldstino. However, their relation becomes evident by going to a different field basis. Let us go to a basis in which the six moduli are given by the sgoldstino and five orthogonal modes, where we take all moduli to be canonically normalized and rotate the orthogonal fields such that their mass submatrix is diagonalized. The full mass matrix is then
\begin{equation} \label{massmatrix-IIB}
V_{ij} = \begin{pmatrix}
  \scriptstyle{7.6 \cdot 10^2} & \scriptstyle{4.5\cdot 10^3} & \scriptstyle{2.8\cdot 10^2} & \scriptstyle{-1.3\cdot 10^4} & \scriptstyle{-34} & \scriptstyle{2.9\cdot 10^4} \\
   \scriptstyle{4.5\cdot 10^3} & \scriptstyle{3.3\cdot 10^5} & 0 & 0 & 0 & 0 \\
   \scriptstyle{2.8\cdot 10^2} & 0 & \scriptstyle{6.2\cdot 10^2} & 0 & 0 & 0 \\
   \scriptstyle{-1.3\cdot 10^4} & 0 & 0 & \scriptstyle{2.1\cdot 10^6} & 0 & 0 \\
   \scriptstyle{-34} & 0 & 0 & 0 & \scriptstyle{1.4\cdot 10^2} & 0 \\
   \scriptstyle{2.9\cdot 10^4} & 0 & 0 & 0 & 0 & \scriptstyle{1.3\cdot 10^6}
\end{pmatrix},
\end{equation}
where the first row contains the sgoldstino mass term $m_\phi^2$ and the off-diagonal components $V_{\phi j}$. In the new basis, the sgoldstino thus corresponds to the vector $\left(1,0,0,0,0,0\right)$. We also observe that three of the orthogonal moduli have very large masses of the order $\mathcal{O}(10^5-10^6)$, while the other two are rather light. The tachyon is given by the eigenvector for the smallest eigenvalue of \eqref{massmatrix-IIB}, which yields
\begin{equation}
\left(0.93, -0.013, -0.35, 0.0059, 0.11, -0.021\right).
\end{equation}
Hence, as claimed, the tachyon predominantly points into the sgoldstino direction but is also rotated slightly into the directions of the orthogonal moduli. One observes that its components along the three heavy directions are extremely small while the mixing with the lighter moduli is stronger. In view of our discussion around \eqref{sgoldstino-mixing}, this is precisely what we should have expected. Recall that there we found that the deviation between the tachyon and the sgoldstino is suppressed by the mass scale of the moduli with which the sgoldstino mixes.

Finally, let us address the observation of \cite{Danielsson:2012et} that, in the class of type IIA models they considered, the ubiquitous tachyons always lie in the moduli subspace spanned by the dilaton, the 6D volume and the O-plane volumes. To this end, we consider the basis
\begin{equation}
\rho_1=\sigma_1^2\sigma_3\sigma_5, \quad \rho_2=\sigma_1^2\sigma_4\sigma_6, \quad \rho_3=\sigma_3\sigma_4, \quad \rho_4=\sigma_5\sigma_6, \quad  \rho_5=\frac{\sigma_3\sigma_6}{\sigma_4\sigma_5}.
\end{equation}
Here, $\rho_1$ and $\rho_2$ are the volume moduli of the O7-planes, $\rho_3$ and $\rho_4$ are the volume moduli of the O5-planes, and $\rho_5$ is an orthogonal mode whose excitation does not change any O-plane volumes. Note that $\text{vol}_6 \sim (\rho_1\rho_2\rho_3\rho_4)^{1/4}$ such that we could make a basis change and trade one of the O-plane volumes for the 6D volume. It is now straightforward to check that the conjecture of \cite{Danielsson:2012et} is indeed true in this model. A simple example for a direction with a negative minor at the critical point \eqref{numerical-dS} is $\sigma_1$, which has $m_{\sigma_1}^2 = -1.5\cdot 10^2$. In the above basis, it is given by $\sigma_1=(\rho_1\rho_2/\rho_3\rho_4)^{1/4}$ and, hence, indeed a combination of the O-plane volumes. It is not clear to us whether this is a coincidence at this dS critical point or whether there is a deeper reason for this. In any case, it certainly shows that the idea of \cite{Danielsson:2012et} can be useful to detect instabilities beyond the class of models considered there.

\section{Conclusions}
\label{concl}

In this note, we gave a simple explanation for the fact that the classical dS extrema known in the literature are unstable, employing a combination of 10D and 4D reasoning. From the 10D perspective, we argued that a key to understand the tachyons is the well-known fact that classical dS vacua require a negatively curved internal space. The group spaces on which the models we considered are compactified can only accommodate this requirement if the internal metric is strongly deformed, which drives all dS critical points close to a singular point in moduli space at which the SUSY breaking scale is small compared to the moduli masses.
The instabilities can then be understood from the perspective of 4D supergravity. In \cite{Covi:2008ea}, the important observation was made that the stability of a dS solution is determined by the sgoldstino mass $m_{\phi}^2$ if the superpotential can be tuned such that all other moduli masses are large compared to the SUSY breaking scale.
Our explicit analysis shows that in general also the off-diagonal terms $V_{\phi j}$ in the mass matrix are crucial for stability. In particular, the tachyons at the classical dS extrema known in the literature do typically not align with the sgoldstino because the $V_{\phi j}$ are not small in these models. Generalizing the sgoldstino no-go theorem to models with non-diagonal mass matrices thus leads to the prediction that tachyons should be expected close to but not exactly aligned with the sgoldstino direction. Assuming large masses of the orthogonal moduli, a sufficient criterion for stability is that the sgoldstino itself is stable, $m_{\phi}^2>0$, and that the off-diagonal terms $V_{\phi j}$ are small or absent. We stress that this observation is not tied to the context of classical dS vacua but may be a useful stability criterion for dS vacua in string theory and supergravity in general.

Our results are consistent with the earlier meta-stability analyses \cite{Covi:2008ea, Danielsson:2012et} and go beyond them in several ways. First, we were able to analytically identify the ubiquitous tachyons in the type IIA solutions we considered and give a satisfactory, general explanation for their appearance.
We stress that our argument holds for dS solutions with finite curvature, whereas previous analytic results in \cite{Danielsson:2012et} applied to the limit infinitesimally close to a Minkowski point. Second, we expect that our arguments are model-independent since no reference to a special set of ingredients is made to explain the origin of the tachyons. We verified this by succesfully identifying the tachyon in a model in type IIB for which no explanation had been available before. It would be interesting to check explicitly whether our argument also works for other examples, e.g., in \cite{Flauger:2008ad, Danielsson:2011au}. Although we believe that the ``fundamental'' explanation for the tachyons is the sgoldstino argument, it is interesting that we also found that the tachyon in the type IIB model lies in the moduli subspace spanned by the O5/O7-plane volumes. This confirms that the O-plane volume moduli advertized in \cite{Danielsson:2012et} can serve as useful proxies to check for tachyons also in type IIB compactifications. It would be interesting to see whether the two viewpoints are related more generally in string compactifications.

The most interesting question is obviously what lessons can be learned from our insights, i.e., how to evade the appearance of tachyons in future searches for dS vacua.
It is important to stress that our arguments do not exclude that stable solutions with a positive cosmological constant or solutions admitting slow-roll inflation with sufficiently many e-folds are still hiding somewhere in the classical landscape of type II string theory. The claim is rather that, \emph{if} any tachyons are present in a given model, they should be expected to appear in the vicinity of the sgoldstino.

An obvious way to get around the constraints we found is to add additional ingredients to the setups such that more terms appear in the scalar potential. This may then relieve the tension between having a positive vacuum energy and a positive definite mass matrix. It is known, for example, that turning on non-geometric fluxes facilitates moduli stabilization \cite{deCarlos:2009fq, deCarlos:2009qm, Dibitetto:2010rg, Danielsson:2012by, Blaback:2013ht, Damian:2013dq, Damian:2013dwa, Hassler:2014mla}. Other examples of possible extra ingredients are less conventional branes, perturbative or non-perturbative quantum corrections or terms from supercritical string theory \cite{Saltman:2004jh, Dong:2010pm, Dodelson:2013iba}. While considering such ingredients is certainly a legitimate strategy, they inevitably make the constructions more complicated. Furthermore, some of the more ``exotic'' ingredients (such as non-geometric fluxes and certain types of branes) are not yet sufficiently well understood, which makes it less clear how to keep the corresponding solutions under control. Some of the appeal of the original idea of classical dS vacua would therefore be lost. For future work, it would therefore be interesting to determine the \emph{minimally} required ingredients in string compactifications to guarantee meta-stability. Our analysis has shown that the crucial objects to consider for such a task are $m_{\phi}^2$ and  $V_{\phi j}$.

In view of our results, one might also ask whether meta-stability could be easier to achieve in compactifications in which the internal space is negatively curved without any strong metric deformations.
For SU(3) structure compactifications with O6-planes in type IIA, an extensive scan of models on group and coset spaces was carried out in \cite{Danielsson:2011au}. Surprisingly, the authors found that nilmanifolds, which are necessarily negatively curved (and, hence, do not require any strong metric deformations), do not admit any dS critical points in these models. This may point towards the existence of a no-go theorem involving both slow-roll parameters such that at least one of the two is always incompatible with dS vacua. It would be interesting to check whether such issues can be avoided in other classes of models. It would also be interesting to consider internal spaces which are not groups or cosets and see whether they can evade the appearance of tachyons. Finally, it would be interesting to explore possible connections of our results to \cite{Kallosh:2014oja, Marsh:2014nla}, where sufficient conditions for meta-stability were found in the context of no-scale models in 4D $\mathcal{N}=1$ supergravity.
We hope to come back to these questions in future work.

\section*{Acknowledgements}

I would like to thank Gautier Solard, Timm Wrase, Thomas Van Riet and Marco Zagermann for useful discussions and correspondence. This work was supported by the DFG Transregional Collaborative Research Centre TRR 33 ``The Dark Universe''. Parts of this work were completed at the HKUST Jockey Club Institute for Advanced Study and supported there by the HKRGC grants 604213 and HKUST4/CRF/13G.

\appendix

\section{A no-go theorem against classical dS vacua?}
\label{app}

Here, we review a no-go theorem formulated in \cite{Dasgupta:2014pma}, where it was argued that dS solutions are excluded in type IIB string theory at the classical level.
In order to show that, the authors of \cite{Dasgupta:2014pma} analyze the 10D Einstein equations and make use of the fact that the negative tension of the O-planes is localized to a certain submanifold of the internal space. Their arguments therefore only apply to solutions with localized O-planes but not to smeared solutions. However, as we will explain below, the theorem can also be evaded in the presence of localized O-planes. It is therefore not a serious obstacle to the construction of classical dS vacua.

We consider a compactification of type IIB supergravity with the usual RR and NSNS field strengths as well as \mbox{(anti-)}D$p$-branes and (anti-)O$p$-planes of arbitrary dimension $p \ge 3$. The 10D spacetime is taken to be the warped product of a 4D external spacetime and a 6D compact space,
\begin{equation}
\d s_{10}^2 = \e^{2A} \tilde g_{\mu\nu}\d x^\mu \d x^\nu + \d s_{6}^2.
\end{equation}
The trace of the external components of the Einstein equations then takes the form
\begin{equation}
\label{einstein}
R^{(4)} = \frac{1}{4}(T_\mu^\mu-T_m^m)^\text{bulk} + \frac{1}{4}(T_\mu^\mu-T_m^m)^\text{loc},
\end{equation}
where we have set $2\kappa_{10}^2=1$. Furthermore, $T_{MN}^\text{bulk}$ denotes the stress tensor associated to the bulk fields and $T_{MN}^\text{loc}$ the one containing the localized sources. One can show that $(T_\mu^\mu-T_m^m)^\text{bulk}$ is always negative, whereas $(T_\mu^\mu-T_m^m)^\text{loc}$ is negative for (anti-)D$p$-branes but positive for (anti-)O$p$-planes. Integrating the above equation over the compact space, one therefore concludes that neither Minkowski nor dS solutions are possible in the absence of O-planes. This is essentially the Maldacena-Nu{\~{n}}ez no-go theorem \cite{Maldacena:2000mw}.

In \cite{Dasgupta:2014pma}, \eqref{einstein} is instead considered pointwise in the internal space. Evaluating it away from the positions of the O-planes, one then observes that their negative tension does not contribute to the stress-energy. The right-hand side is therefore manifestly negative and, hence, $R^{(4)} < 0$. The authors of \cite{Dasgupta:2014pma} take this to conclude that there are no classical dS vacua in type IIB string theory if the spacetime is a direct product, i.e., in the absence of warping. However, flux backgrounds with localized sources are generically warped (except in the special case where all tadpoles are cancelled locally). Moreover, warping corrections to the equations of motion are large everywhere on the internal space in such compactifications, as was first pointed out in \cite{Douglas:2010rt} and later verified for explicit examples in \cite{Blaback:2010sj}.\footnote{By ``large'' we mean here that the warping terms are nowhere subleading compared to the other terms in the 10D equations of motion. A detailed discussion of this point can also be found in Section 2.4.2 of \cite{Junghans:2013xza}.}

In order to determine the 4D vacuum energy, we have to compute the external Ricci scalar in 4D Einstein frame. In a warped background, the relevant object to consider is then not $R^{(4)}$ but (up to a constant volume factor) the Ricci scalar of the unwarped external metric $\tilde R^{(4)}$, where
\begin{equation}
R^{(4)} = \e^{-2A}\tilde R^{(4)} - \e^{-4A} \nabla^2 \e^{4A}.
\end{equation}
Substituting this into \eqref{einstein}, we find that the sign of $\tilde R^{(4)}$ is undetermined for a general matter content and depends on the sign and magnitude of the warping term. The sign of $R^{(4)}$, on the other hand, is not related to the sign of the cosmological constant unless warping effects are negligible. Hence, the Einstein equations are not sufficient to exclude classical dS vacua in the presence of localized O-planes. If one instead considers a solution with smeared sources, there is usually no warping. In that case, however, the negative tension of the O-planes contributes everywhere on the compact space such that dS solutions are again not excluded by the Einstein equations.

In order to address warping effects, a second argument is presented in \cite{Dasgupta:2014pma}, which again makes use of the Einstein equations. Instead of considering \eqref{einstein} pointwise in the internal space, the authors propose to integrate the equation over the internal manifold up to a boundary which they choose such that the locations of the O-planes are not contained in the integrated region. Let us denote the integrated region by $\mathcal{M}_1$, the excluded region containing the O-planes by $\mathcal{M}_2$ and the boundary between the two regions by $\partial\mathcal{M}$. One then finds
\begin{equation}
\label{einstein-boundary}
\int_{\mathcal{M}_1}\!\!\! \d^6 x \sqrt{g_6}\, \e^{2A} \tilde R^{(4)} = -\oint_{\partial\mathcal{M}}\!\!\! \star_6 \d \e^{4A} + \int_{\mathcal{M}_1}\!\!\! \d^6 x \sqrt{g_6}\, \frac{1}{4}\e^{4A} \left[ (T_\mu^\mu-T_m^m)^\text{bulk} + (T_\mu^\mu-T_m^m)^\text{branes} \right].
\end{equation}
It is then pointed out in \cite{Dasgupta:2014pma} that, assuming the boundary term on the right-hand side is zero, one has $\tilde R^{(4)} < 0$ since both the bulk contribution and the (anti-)brane contribution to the right-hand side are negative. However, the boundary term is in general not zero but can become positive if the O-planes that sit inside of the excluded region carry enough negative tension.\footnote{This was also pointed out in the revised version of \cite{Dasgupta:2014pma}.} It is straightforward to verify this by looking at explicit examples of flux compactifications with localized O-planes such as those in \cite{Giddings:2001yu}, which indeed yields
\begin{equation}
-\oint_{\partial\mathcal{M}}\!\!\!  \star_6 \d \e^{4A} > 0.
\end{equation}
In the solutions of \cite{Giddings:2001yu}, the value of the boundary term is such that it exactly cancels the negative second term on the right-hand side of \eqref{einstein-boundary}. Hence, the solutions describe Minkowski vacua, $\tilde R^{(4)}=0$. For a general solution, however, the magnitude of the boundary term is a priori undetermined and, hence, $\tilde R^{(4)}$ can in general be positive, zero or negative. 

It is furthermore argued in \cite{Dasgupta:2014pma} that, in order to resolve the O-plane singularities, string corrections are necessarily large such that classical dS vacua are not under control in the presence of O-planes. However, in the usual limit of large volume and small string coupling, these corrections are expected to be suppressed everywhere except very close to the sources such that their contribution to the 4D vacuum energy is negligible, as utilized in basically every moduli stabilization scenario.
We therefore conclude that the no-go theorem of \cite{Dasgupta:2014pma} does not pose stronger restrictions on the existence of Minkowski and dS vacua than the theorem of \cite{Maldacena:2000mw}: in order to obtain Minkowski or dS solutions at the classical level, O-planes are required to be present.
We should stress that this is only a necessary condition and does of course not guarantee that classical dS vacua must exist. As we discussed in this paper, there are several open problems related to this question such that the explicit construction of such solutions remains a challenge for future research.

\bibliographystyle{utphys}
\bibliography{groups}

\providecommand{\href}[2]{#2}\begingroup\raggedright\begin{thebibliography}{10}

\bibitem{Cicoli:2013cha}
M.~Cicoli, D.~Klevers, S.~Krippendorf, C.~Mayrhofer, F.~Quevedo and
  R.~Valandro,  {\em {Explicit de Sitter Flux Vacua for Global String Models
  with Chiral Matter}}, JHEP {\bf 1405} (2014) 001
[\href{http://www.arXiv.org/abs/1312.0014}{{\tt 1312.0014}}].
%%CITATION = ARXIV:1312.0014;%%.

\bibitem{Gibbons:1984kp}
G.~W. Gibbons,  {\em {Aspects of Supergravity Theories}}, in {\em
  Supersymmetry, Supergravity and Related Topics}, F.~del Aguila, J.~A.
  de~Azc{\'{a}}rraga and L.~E. Ib{\'{a}}{\~{n}}ez, eds., pp.~346--351.
\newblock World Scientific, 1985.

\bibitem{deWit:1986xg}
B.~de~Wit, D.~J. Smit and N.~D. Hari~Dass,  {\em {Residual Supersymmetry of
  Compactified D=10 Supergravity}}, Nucl.Phys. {\bf B283} (1987)
165.
%%CITATION = NUPHA,B283,165;%%.

\bibitem{Maldacena:2000mw}
J.~M. Maldacena and C.~Nu{\~{n}}ez,  {\em {Supergravity description of field
  theories on curved manifolds and a no go theorem}}, Int.J.Mod.Phys. {\bf A16}
  (2001) 822--855
[\href{http://www.arXiv.org/abs/hep-th/0007018}{{\tt hep-th/0007018}}].
%%CITATION = HEP-TH/0007018;%%.

\bibitem{Koerber:2007xk}
P.~Koerber and L.~Martucci,  {\em {From ten to four and back again: how to
  generalize the geometry}}, JHEP {\bf 08} (2007) 059
[\href{http://www.arXiv.org/abs/0707.1038}{{\tt 0707.1038}}].
%%CITATION = 0707.1038;%%.

\bibitem{Baumann:2010sx}
D.~Baumann, A.~Dymarsky, S.~Kachru, I.~R. Klebanov and L.~McAllister,  {\em
  {D3-brane Potentials from Fluxes in AdS/CFT}}, JHEP {\bf 1006} (2010) 072
[\href{http://www.arXiv.org/abs/1001.5028}{{\tt 1001.5028}}].
%%CITATION = ARXIV:1001.5028;%%.

\bibitem{Heidenreich:2010ad}
B.~Heidenreich, L.~McAllister and G.~Torroba,  {\em {Dynamic SU(2) Structure
  from Seven-branes}}, JHEP {\bf 1105} (2011) 110
[\href{http://www.arXiv.org/abs/1011.3510}{{\tt 1011.3510}}].
%%CITATION = ARXIV:1011.3510;%%.

\bibitem{Dymarsky:2010mf}
A.~Dymarsky and L.~Martucci,  {\em {D-brane non-perturbative effects and
  geometric deformations}}, JHEP {\bf 1104} (2011) 061
[\href{http://www.arXiv.org/abs/1012.4018}{{\tt 1012.4018}}].
%%CITATION = ARXIV:1012.4018;%%.

\bibitem{Kachru:2003aw}
S.~Kachru, R.~Kallosh, A.~D. Linde and S.~P. Trivedi,  {\em {De Sitter vacua in
  string theory}}, Phys.Rev. {\bf D68} (2003) 046005
[\href{http://www.arXiv.org/abs/hep-th/0301240}{{\tt hep-th/0301240}}].
%%CITATION = HEP-TH/0301240;%%.

\bibitem{Bergshoeff:2015jxa}
E.~A. Bergshoeff, K.~Dasgupta, R.~Kallosh, A.~Van~Proeyen and T.~Wrase,  {\em
  {$ \overline{\mathrm{D}3} $ and dS}}, JHEP {\bf 1505} (2015) 058
[\href{http://www.arXiv.org/abs/1502.07627}{{\tt 1502.07627}}].
%%CITATION = ARXIV:1502.07627;%%.

\bibitem{Balasubramanian:2005zx}
V.~Balasubramanian, P.~Berglund, J.~P. Conlon and F.~Quevedo,  {\em
  {Systematics of moduli stabilisation in Calabi-Yau flux compactifications}},
  JHEP {\bf 0503} (2005) 007
[\href{http://www.arXiv.org/abs/hep-th/0502058}{{\tt hep-th/0502058}}].
%%CITATION = HEP-TH/0502058;%%.

\bibitem{Conlon:2005ki}
J.~P. Conlon, F.~Quevedo and K.~Suruliz,  {\em {Large-volume flux
  compactifications: Moduli spectrum and D3/D7 soft supersymmetry breaking}},
  JHEP {\bf 0508} (2005) 007
[\href{http://www.arXiv.org/abs/hep-th/0505076}{{\tt hep-th/0505076}}].
%%CITATION = HEP-TH/0505076;%%.

\bibitem{Balasubramanian:2004uy}
V.~Balasubramanian and P.~Berglund,  {\em {Stringy corrections to K{\"{a}}hler
  potentials, SUSY breaking, and the cosmological constant problem}}, JHEP {\bf
  0411} (2004) 085
[\href{http://www.arXiv.org/abs/hep-th/0408054}{{\tt hep-th/0408054}}].
%%CITATION = HEP-TH/0408054;%%.

\bibitem{Westphal:2006tn}
A.~Westphal,  {\em {de Sitter string vacua from K{\"{a}}hler uplifting}}, JHEP
  {\bf 0703} (2007) 102
[\href{http://www.arXiv.org/abs/hep-th/0611332}{{\tt hep-th/0611332}}].
%%CITATION = HEP-TH/0611332;%%.

\bibitem{Rummel:2011cd}
M.~Rummel and A.~Westphal,  {\em {A sufficient condition for de Sitter vacua in
  type IIB string theory}}, JHEP {\bf 1201} (2012) 020
[\href{http://www.arXiv.org/abs/1107.2115}{{\tt 1107.2115}}].
%%CITATION = ARXIV:1107.2115;%%.

\bibitem{Louis:2012nb}
J.~Louis, M.~Rummel, R.~Valandro and A.~Westphal,  {\em {Building an explicit
  de Sitter}}, JHEP {\bf 1210} (2012) 163
[\href{http://www.arXiv.org/abs/1208.3208}{{\tt 1208.3208}}].
%%CITATION = ARXIV:1208.3208;%%.

\bibitem{Cicoli:2012fh}
M.~Cicoli, A.~Maharana, F.~Quevedo and C.~P. Burgess,  {\em {De Sitter String
  Vacua from Dilaton-dependent Non-perturbative Effects}}, JHEP {\bf 1206}
  (2012) 011
[\href{http://www.arXiv.org/abs/1203.1750}{{\tt 1203.1750}}].
%%CITATION = ARXIV:1203.1750;%%.

\bibitem{Blaback:2013qza}
J.~Bl{\r{a}}b{\"{a}}ck, D.~Roest and I.~Zavala,  {\em {De Sitter Vacua from
  Non-perturbative Flux Compactifications}}, Phys.Rev. {\bf D90} (2014) 024065
[\href{http://www.arXiv.org/abs/1312.5328}{{\tt 1312.5328}}].
%%CITATION = ARXIV:1312.5328;%%.

\bibitem{Danielsson:2013rza}
U.~Danielsson and G.~Dibitetto,  {\em {An alternative to anti-branes and
  O-planes?}}, JHEP {\bf 1405} (2014) 013
[\href{http://www.arXiv.org/abs/1312.5331}{{\tt 1312.5331}}].
%%CITATION = ARXIV:1312.5331;%%.

\bibitem{Rummel:2014raa}
M.~Rummel and Y.~Sumitomo,  {\em {De Sitter Vacua from a D-term Generated
  Racetrack Uplift}}, JHEP {\bf 1501} (2015) 015
[\href{http://www.arXiv.org/abs/1407.7580}{{\tt 1407.7580}}].
%%CITATION = ARXIV:1407.7580;%%.

\bibitem{Kallosh:2014oja}
R.~Kallosh, A.~Linde, B.~Vercnocke and T.~Wrase,  {\em {Analytic Classes of
  Metastable de Sitter Vacua}}, JHEP {\bf 10} (2014) 11
[\href{http://www.arXiv.org/abs/1406.4866}{{\tt 1406.4866}}].
%%CITATION = ARXIV:1406.4866;%%.

\bibitem{Marsh:2014nla}
M.~C.~D. Marsh, B.~Vercnocke and T.~Wrase,  {\em {Decoupling and de Sitter
  Vacua in Approximate No-Scale Supergravities}}, JHEP {\bf 05} (2015) 081
[\href{http://www.arXiv.org/abs/1411.6625}{{\tt 1411.6625}}].
%%CITATION = ARXIV:1411.6625;%%.

\bibitem{Guarino:2015gos}
A.~Guarino and G.~Inverso,  {\em {Single-step de Sitter vacua from
  non-perturbative effects with matter}},
\href{http://www.arXiv.org/abs/1511.07841}{{\tt 1511.07841}}.
%%CITATION = ARXIV:1511.07841;%%.

\bibitem{Retolaza:2015nvh}
A.~Retolaza and A.~Uranga,  {\em {De Sitter Uplift with Dynamical Susy
  Breaking}},
\href{http://www.arXiv.org/abs/1512.06363}{{\tt 1512.06363}}.
%%CITATION = ARXIV:1512.06363;%%.

\bibitem{deCarlos:2009fq}
B.~de~Carlos, A.~Guarino and J.~M. Moreno,  {\em {Flux moduli stabilisation,
  Supergravity algebras and no-go theorems}}, JHEP {\bf 1001} (2010) 012
[\href{http://www.arXiv.org/abs/0907.5580}{{\tt 0907.5580}}].
%%CITATION = ARXIV:0907.5580;%%.

\bibitem{deCarlos:2009qm}
B.~de~Carlos, A.~Guarino and J.~M. Moreno,  {\em {Complete classification of
  Minkowski vacua in generalised flux models}}, JHEP {\bf 1002} (2010) 076
[\href{http://www.arXiv.org/abs/0911.2876}{{\tt 0911.2876}}].
%%CITATION = ARXIV:0911.2876;%%.

\bibitem{Dibitetto:2010rg}
G.~Dibitetto, R.~Linares and D.~Roest,  {\em {Flux Compactifications, Gauge
  Algebras and De Sitter}}, Phys.Lett. {\bf B688} (2010) 96--100
[\href{http://www.arXiv.org/abs/1001.3982}{{\tt 1001.3982}}].
%%CITATION = ARXIV:1001.3982;%%.

\bibitem{Danielsson:2012by}
U.~Danielsson and G.~Dibitetto,  {\em {On the distribution of stable de Sitter
  vacua}}, JHEP {\bf 03} (2013) 018
[\href{http://www.arXiv.org/abs/1212.4984}{{\tt 1212.4984}}].
%%CITATION = ARXIV:1212.4984;%%.

\bibitem{Blaback:2013ht}
J.~Bl{\r{a}}b{\"{a}}ck, U.~Danielsson and G.~Dibitetto,  {\em {Fully stable dS
  vacua from generalised fluxes}}, JHEP {\bf 08} (2013) 054
[\href{http://www.arXiv.org/abs/1301.7073}{{\tt 1301.7073}}].
%%CITATION = ARXIV:1301.7073;%%.

\bibitem{Damian:2013dq}
C.~Damian, L.~R. D{\'{i}}az-Barr{\'{o}}n, O.~Loaiza-Brito and M.~Sabido,  {\em
  {Slow-Roll Inflation in Non-geometric Flux Compactification}}, JHEP {\bf 06}
  (2013) 109
[\href{http://www.arXiv.org/abs/1302.0529}{{\tt 1302.0529}}].
%%CITATION = ARXIV:1302.0529;%%.

\bibitem{Damian:2013dwa}
C.~Damian and O.~Loaiza-Brito,  {\em {More stable de Sitter vacua from S-dual
  nongeometric fluxes}}, Phys. Rev. {\bf D88} (2013), no.~4, 046008
[\href{http://www.arXiv.org/abs/1304.0792}{{\tt 1304.0792}}].
%%CITATION = ARXIV:1304.0792;%%.

\bibitem{Hassler:2014mla}
F.~Hassler, D.~Lust and S.~Massai,  {\em {On Inflation and de Sitter in
  Non-Geometric String Backgrounds}},
\href{http://www.arXiv.org/abs/1405.2325}{{\tt 1405.2325}}.
%%CITATION = ARXIV:1405.2325;%%.

\bibitem{Danielsson:2015tsa}
U.~Danielsson and G.~Dibitetto,  {\em {Geometric non-geometry}}, JHEP {\bf 04}
  (2015) 084
[\href{http://www.arXiv.org/abs/1501.03944}{{\tt 1501.03944}}].
%%CITATION = ARXIV:1501.03944;%%.

\bibitem{Blumenhagen:2015zma}
R.~Blumenhagen, P.~d. Bosque, F.~Hassler and D.~Lust,  {\em {Generalized Metric
  Formulation of Double Field Theory on Group Manifolds}}, JHEP {\bf 08} (2015)
  056
[\href{http://www.arXiv.org/abs/1502.02428}{{\tt 1502.02428}}].
%%CITATION = ARXIV:1502.02428;%%.

\bibitem{Blumenhagen:2015lta}
R.~Blumenhagen, A.~Font and E.~Plauschinn,  {\em {Relating Double Field Theory
  to the Scalar Potential of N=2 Gauged Supergravity}},
\href{http://www.arXiv.org/abs/1507.08059}{{\tt 1507.08059}}.
%%CITATION = ARXIV:1507.08059;%%.

\bibitem{Hertzberg:2007wc}
M.~P. Hertzberg, S.~Kachru, W.~Taylor and M.~Tegmark,  {\em {Inflationary
  Constraints on Type IIA String Theory}}, JHEP {\bf 0712} (2007) 095
[\href{http://www.arXiv.org/abs/0711.2512}{{\tt 0711.2512}}].
%%CITATION = ARXIV:0711.2512;%%.

\bibitem{Silverstein:2007ac}
E.~Silverstein,  {\em {Simple de Sitter Solutions}}, Phys.Rev. {\bf D77} (2008)
  106006
[\href{http://www.arXiv.org/abs/0712.1196}{{\tt 0712.1196}}].
%%CITATION = ARXIV:0712.1196;%%.

\bibitem{Haque:2008jz}
S.~S. Haque, G.~Shiu, B.~Underwood and T.~Van~Riet,  {\em {Minimal simple de
  Sitter solutions}}, Phys.Rev. {\bf D79} (2009) 086005
[\href{http://www.arXiv.org/abs/0810.5328}{{\tt 0810.5328}}].
%%CITATION = ARXIV:0810.5328;%%.

\bibitem{Steinhardt:2008nk}
P.~J. Steinhardt and D.~Wesley,  {\em {Dark Energy, Inflation and Extra
  Dimensions}}, Phys.Rev. {\bf D79} (2009) 104026
[\href{http://www.arXiv.org/abs/0811.1614}{{\tt 0811.1614}}].
%%CITATION = ARXIV:0811.1614;%%.

\bibitem{Caviezel:2008tf}
C.~Caviezel, P.~Koerber, S.~K{\"{o}}rs, D.~Lust, T.~Wrase and M.~Zagermann,
  {\em {On the Cosmology of Type IIA Compactifications on SU(3)-structure
  Manifolds}}, JHEP {\bf 0904} (2009) 010
[\href{http://www.arXiv.org/abs/0812.3551}{{\tt 0812.3551}}].
%%CITATION = ARXIV:0812.3551;%%.

\bibitem{Flauger:2008ad}
R.~Flauger, S.~Paban, D.~Robbins and T.~Wrase,  {\em {Searching for slow-roll
  moduli inflation in massive type IIA supergravity with metric fluxes}},
  Phys.Rev. {\bf D79} (2009) 086011
[\href{http://www.arXiv.org/abs/0812.3886}{{\tt 0812.3886}}].
%%CITATION = ARXIV:0812.3886;%%.

\bibitem{Danielsson:2009ff}
U.~H. Danielsson, S.~S. Haque, G.~Shiu and T.~Van~Riet,  {\em {Towards
  Classical de Sitter Solutions in String Theory}}, JHEP {\bf 0909} (2009) 114
[\href{http://www.arXiv.org/abs/0907.2041}{{\tt 0907.2041}}].
%%CITATION = ARXIV:0907.2041;%%.

\bibitem{Caviezel:2009tu}
C.~Caviezel, T.~Wrase and M.~Zagermann,  {\em {Moduli Stabilization and
  Cosmology of Type IIB on SU(2)-Structure Orientifolds}}, JHEP {\bf 1004}
  (2010) 011
[\href{http://www.arXiv.org/abs/0912.3287}{{\tt 0912.3287}}].
%%CITATION = ARXIV:0912.3287;%%.

\bibitem{Wrase:2010ew}
T.~Wrase and M.~Zagermann,  {\em {On Classical de Sitter Vacua in String
  Theory}}, Fortsch.Phys. {\bf 58} (2010) 906--910
[\href{http://www.arXiv.org/abs/1003.0029}{{\tt 1003.0029}}].
%%CITATION = ARXIV:1003.0029;%%.

\bibitem{VanRiet:2011yc}
T.~Van~Riet,  {\em {On classical de Sitter solutions in higher dimensions}},
  Class.Quant.Grav. {\bf 29} (2012) 055001
[\href{http://www.arXiv.org/abs/1111.3154}{{\tt 1111.3154}}].
%%CITATION = ARXIV:1111.3154;%%.

\bibitem{Gautason:2012tb}
F.~F. Gautason, D.~Junghans and M.~Zagermann,  {\em {On Cosmological Constants
  from alpha'-Corrections}}, JHEP {\bf 1206} (2012) 029
[\href{http://www.arXiv.org/abs/1204.0807}{{\tt 1204.0807}}].
%%CITATION = ARXIV:1204.0807;%%.

\bibitem{Green:2011cn}
S.~R. Green, E.~J. Martinec, C.~Quigley and S.~Sethi,  {\em {Constraints on
  String Cosmology}}, Class. Quant. Grav. {\bf 29} (2012) 075006
[\href{http://www.arXiv.org/abs/1110.0545}{{\tt 1110.0545}}].
%%CITATION = ARXIV:1110.0545;%%.

\bibitem{Kutasov:2015eba}
D.~Kutasov, T.~Maxfield, I.~Melnikov and S.~Sethi,  {\em {Constraining de
  Sitter Space in String Theory}}, Phys. Rev. Lett. {\bf 115} (2015), no.~7,
  071305
[\href{http://www.arXiv.org/abs/1504.00056}{{\tt 1504.00056}}].
%%CITATION = ARXIV:1504.00056;%%.

\bibitem{Quigley:2015jia}
C.~Quigley,  {\em {Gaugino Condensation and the Cosmological Constant}}, JHEP
  {\bf 1506} (2015) 104
[\href{http://www.arXiv.org/abs/1504.00652}{{\tt 1504.00652}}].
%%CITATION = ARXIV:1504.00652;%%.

\bibitem{Andriot:2010ju}
D.~Andriot, E.~Goi, R.~Minasian and M.~Petrini,  {\em {Supersymmetry breaking
  branes on solvmanifolds and de Sitter vacua in string theory}}, JHEP {\bf
  1105} (2011) 028
[\href{http://www.arXiv.org/abs/1003.3774}{{\tt 1003.3774}}].
%%CITATION = ARXIV:1003.3774;%%.

\bibitem{Saltman:2004jh}
A.~Saltman and E.~Silverstein,  {\em {A new handle on de Sitter
  compactifications}}, JHEP {\bf 0601} (2006) 139
[\href{http://www.arXiv.org/abs/hep-th/0411271}{{\tt hep-th/0411271}}].
%%CITATION = HEP-TH/0411271;%%.

\bibitem{Dong:2010pm}
X.~Dong, B.~Horn, E.~Silverstein and G.~Torroba,  {\em {Micromanaging de Sitter
  holography}}, Class.Quant.Grav. {\bf 27} (2010) 245020
[\href{http://www.arXiv.org/abs/1005.5403}{{\tt 1005.5403}}].
%%CITATION = ARXIV:1005.5403;%%.

\bibitem{Dodelson:2013iba}
M.~Dodelson, X.~Dong, E.~Silverstein and G.~Torroba,  {\em {New solutions with
  accelerated expansion in string theory}}, JHEP {\bf 12} (2014) 050
[\href{http://www.arXiv.org/abs/1310.5297}{{\tt 1310.5297}}].
%%CITATION = ARXIV:1310.5297;%%.

\bibitem{Dasgupta:2014pma}
K.~Dasgupta, R.~Gwyn, E.~McDonough, M.~Mia and R.~Tatar,  {\em {de Sitter Vacua
  in Type IIB String Theory: Classical Solutions and Quantum Corrections}},
  JHEP {\bf 1407} (2014) 054
[\href{http://www.arXiv.org/abs/1402.5112}{{\tt 1402.5112}}].
%%CITATION = ARXIV:1402.5112;%%.

\bibitem{Danielsson:2010bc}
U.~H. Danielsson, P.~Koerber and T.~Van~Riet,  {\em {Universal de Sitter
  solutions at tree-level}}, JHEP {\bf 1005} (2010) 090
[\href{http://www.arXiv.org/abs/1003.3590}{{\tt 1003.3590}}].
%%CITATION = ARXIV:1003.3590;%%.

\bibitem{Danielsson:2011au}
U.~H. Danielsson, S.~S. Haque, P.~Koerber, G.~Shiu, T.~Van~Riet and T.~Wrase,
  {\em {De Sitter hunting in a classical landscape}}, Fortsch.Phys. {\bf 59}
  (2011) 897--933
[\href{http://www.arXiv.org/abs/1103.4858}{{\tt 1103.4858}}].
%%CITATION = ARXIV:1103.4858;%%.

\bibitem{Gur-Ari:2013sba}
G.~Gur-Ari,  {\em {Brane Inflation and Moduli Stabilization on Twisted Tori}},
  JHEP {\bf 01} (2014) 179
[\href{http://www.arXiv.org/abs/1310.6787}{{\tt 1310.6787}}].
%%CITATION = ARXIV:1310.6787;%%.

\bibitem{Blaback:2013fca}
J.~Bl{\r{a}}b{\"{a}}ck, U.~Danielsson and G.~Dibitetto,  {\em {Accelerated
  Universes from type IIA Compactifications}}, JCAP {\bf 1403} (2014) 003
[\href{http://www.arXiv.org/abs/1310.8300}{{\tt 1310.8300}}].
%%CITATION = ARXIV:1310.8300;%%.

\bibitem{Covi:2008ea}
L.~Covi, M.~G{\'{o}}mez-Reino, C.~Gross, J.~Louis, G.~A. Palma and C.~A.
  Scrucca,  {\em {de Sitter vacua in no-scale supergravities and Calabi-Yau
  string models}}, JHEP {\bf 0806} (2008) 057
[\href{http://www.arXiv.org/abs/0804.1073}{{\tt 0804.1073}}].
%%CITATION = ARXIV:0804.1073;%%.

\bibitem{Shiu:2011zt}
G.~Shiu and Y.~Sumitomo,  {\em {Stability Constraints on Classical de Sitter
  Vacua}}, JHEP {\bf 1109} (2011) 052
[\href{http://www.arXiv.org/abs/1107.2925}{{\tt 1107.2925}}].
%%CITATION = ARXIV:1107.2925;%%.

\bibitem{Danielsson:2012et}
U.~H. Danielsson, G.~Shiu, T.~Van~Riet and T.~Wrase,  {\em {A note on obstinate
  tachyons in classical dS solutions}}, JHEP {\bf 1303} (2013) 138
[\href{http://www.arXiv.org/abs/1212.5178}{{\tt 1212.5178}}].
%%CITATION = ARXIV:1212.5178;%%.

\bibitem{Marsh:2011aa}
D.~Marsh, L.~McAllister and T.~Wrase,  {\em {The Wasteland of Random
  Supergravities}}, JHEP {\bf 1203} (2012) 102
[\href{http://www.arXiv.org/abs/1112.3034}{{\tt 1112.3034}}].
%%CITATION = ARXIV:1112.3034;%%.

\bibitem{Chen:2011ac}
X.~Chen, G.~Shiu, Y.~Sumitomo and S.-H.~H. Tye,  {\em {A Global View on The
  Search for de-Sitter Vacua in (type IIA) String Theory}}, JHEP {\bf 1204}
  (2012) 026
[\href{http://www.arXiv.org/abs/1112.3338}{{\tt 1112.3338}}].
%%CITATION = ARXIV:1112.3338;%%.

\bibitem{Bachlechner:2012at}
T.~C. Bachlechner, D.~Marsh, L.~McAllister and T.~Wrase,  {\em {Supersymmetric
  Vacua in Random Supergravity}}, JHEP {\bf 01} (2013) 136
[\href{http://www.arXiv.org/abs/1207.2763}{{\tt 1207.2763}}].
%%CITATION = ARXIV:1207.2763;%%.

\bibitem{Martucci:2014ska}
L.~Martucci,  {\em {Warping the K{\"{a}}hler potential of F-theory/IIB flux
  compactifications}}, JHEP {\bf 03} (2015) 067
[\href{http://www.arXiv.org/abs/1411.2623}{{\tt 1411.2623}}].
%%CITATION = ARXIV:1411.2623;%%.

\bibitem{Grimm:2014efa}
T.~W. Grimm, T.~G. Pugh and M.~Weissenbacher,  {\em {The effective action of
  warped M-theory reductions with higher derivative terms - Part I}},
\href{http://www.arXiv.org/abs/1412.5073}{{\tt 1412.5073}}.
%%CITATION = ARXIV:1412.5073;%%.

\bibitem{Blaback:2014tfa}
J.~Bl{\r{a}}b{\"{a}}ck, U.~H. Danielsson, D.~Junghans, T.~Van~Riet and S.~C.
  Vargas,  {\em {Localised anti-branes in non-compact throats at zero and
  finite $T$}}, JHEP {\bf 02} (2015) 018
[\href{http://www.arXiv.org/abs/1409.0534}{{\tt 1409.0534}}].
%%CITATION = ARXIV:1409.0534;%%.

\bibitem{Bena:2014jaa}
I.~Bena, M.~Gra{\~{n}}a, S.~Kuperstein and S.~Massai,  {\em {Giant Tachyons in
  the Landscape}}, JHEP {\bf 02} (2015) 146
[\href{http://www.arXiv.org/abs/1410.7776}{{\tt 1410.7776}}].
%%CITATION = ARXIV:1410.7776;%%.

\bibitem{Hartnett:2015oda}
G.~S. Hartnett,  {\em {Localised Anti-Branes in Flux Backgrounds}}, JHEP {\bf
  06} (2015) 007
[\href{http://www.arXiv.org/abs/1501.06568}{{\tt 1501.06568}}].
%%CITATION = ARXIV:1501.06568;%%.

\bibitem{Michel:2014lva}
B.~Michel, E.~Mintun, J.~Polchinski, A.~Puhm and P.~Saad,  {\em {Remarks on
  brane and antibrane dynamics}},
\href{http://www.arXiv.org/abs/1412.5702}{{\tt 1412.5702}}.
%%CITATION = ARXIV:1412.5702;%%.

\bibitem{Blaback:2011nz}
J.~Bl{\r{a}}b{\"{a}}ck, U.~H. Danielsson, D.~Junghans, T.~Van~Riet, T.~Wrase
  and M.~Zagermann,  {\em {The problematic backreaction of SUSY-breaking
  branes}}, JHEP {\bf 1108} (2011) 105
[\href{http://www.arXiv.org/abs/1105.4879}{{\tt 1105.4879}}].
%%CITATION = ARXIV:1105.4879;%%.

\bibitem{Apruzzi:2013yva}
F.~Apruzzi, M.~Fazzi, D.~Rosa and A.~Tomasiello,  {\em {All AdS$_7$ solutions
  of type II supergravity}}, JHEP {\bf 04} (2014) 064
[\href{http://www.arXiv.org/abs/1309.2949}{{\tt 1309.2949}}].
%%CITATION = ARXIV:1309.2949;%%.

\bibitem{Junghans:2014wda}
D.~Junghans, D.~Schmidt and M.~Zagermann,  {\em {Curvature-induced Resolution
  of Anti-brane Singularities}}, JHEP {\bf 10} (2014) 34
[\href{http://www.arXiv.org/abs/1402.6040}{{\tt 1402.6040}}].
%%CITATION = ARXIV:1402.6040;%%.

\bibitem{Danielsson:2013qfa}
U.~H. Danielsson, G.~Dibitetto, M.~Fazzi and T.~Van~Riet,  {\em {A note on
  smeared branes in flux vacua and gauged supergravity}}, JHEP {\bf 1404}
  (2014) 025
[\href{http://www.arXiv.org/abs/1311.6470}{{\tt 1311.6470}}].
%%CITATION = ARXIV:1311.6470;%%.

\bibitem{Banks:2006hg}
T.~Banks and K.~van~den Broek,  {\em {Massive IIA flux compactifications and
  U-dualities}}, JHEP {\bf 0703} (2007) 068
[\href{http://www.arXiv.org/abs/hep-th/0611185}{{\tt hep-th/0611185}}].
%%CITATION = HEP-TH/0611185;%%.

\bibitem{McOrist:2012yc}
J.~McOrist and S.~Sethi,  {\em {M-theory and Type IIA Flux Compactifications}},
  JHEP {\bf 1212} (2012) 122
[\href{http://www.arXiv.org/abs/1208.0261}{{\tt 1208.0261}}].
%%CITATION = ARXIV:1208.0261;%%.

\bibitem{Saracco:2012wc}
F.~Saracco and A.~Tomasiello,  {\em {Localized O6-plane solutions with Romans
  mass}}, JHEP {\bf 1207} (2012) 077
[\href{http://www.arXiv.org/abs/1201.5378}{{\tt 1201.5378}}].
%%CITATION = ARXIV:1201.5378;%%.

\bibitem{Gautason:2015tig}
F.~F. Gautason, M.~Schillo, T.~Van~Riet and M.~Williams,  {\em {Remarks on
  scale separation in flux vacua}},
\href{http://www.arXiv.org/abs/1512.00457}{{\tt 1512.00457}}.
%%CITATION = ARXIV:1512.00457;%%.

\bibitem{Petrini:2013ika}
M.~Petrini, G.~Solard and T.~Van~Riet,  {\em {AdS vacua with scale separation
  from IIB supergravity}}, JHEP {\bf 11} (2013) 010
[\href{http://www.arXiv.org/abs/1308.1265}{{\tt 1308.1265}}].
%%CITATION = ARXIV:1308.1265;%%.

\bibitem{Douglas:2010rt}
M.~R. Douglas and R.~Kallosh,  {\em {Compactification on negatively curved
  manifolds}}, JHEP {\bf 1006} (2010) 004
[\href{http://www.arXiv.org/abs/1001.4008}{{\tt 1001.4008}}].
%%CITATION = ARXIV:1001.4008;%%.

\bibitem{Blaback:2010sj}
J.~Bl{\r{a}}b{\"{a}}ck, U.~H. Danielsson, D.~Junghans, T.~Van~Riet, T.~Wrase
  and M.~Zagermann,  {\em {Smeared versus localised sources in flux
  compactifications}}, JHEP {\bf 1012} (2010) 043
[\href{http://www.arXiv.org/abs/1009.1877}{{\tt 1009.1877}}].
%%CITATION = ARXIV:1009.1877;%%.

\bibitem{Junghans:2013xza}
D.~Junghans,  {\em {Backreaction of Localised Sources in String
  Compactifications}},
\href{http://www.arXiv.org/abs/1309.5990}{{\tt 1309.5990}}.
%%CITATION = ARXIV:1309.5990;%%.

\bibitem{Giddings:2001yu}
S.~B. Giddings, S.~Kachru and J.~Polchinski,  {\em {Hierarchies from fluxes in
  string compactifications}}, Phys.Rev. {\bf D66} (2002) 106006
[\href{http://www.arXiv.org/abs/hep-th/0105097}{{\tt hep-th/0105097}}].
%%CITATION = HEP-TH/0105097;%%.

\end{thebibliography}\endgroup

\end{document}